\documentclass{jfm}
\usepackage{graphicx}
\usepackage{natbib}
\usepackage{amsmath}
\usepackage{color}

\ifCUPmtlplainloaded \else
  \checkfont{eurm10}
  \iffontfound
    \IfFileExists{upmath.sty}
      {\typeout{^^JFound AMS Euler Roman fonts on the system,
                   using the 'upmath' package.^^J}%
       \usepackage{upmath}}
      {\typeout{^^JFound AMS Euler Roman fonts on the system, but you
                   dont seem to have the}%
       \typeout{'upmath' package installed. JFM.cls can take advantage
                 of these fonts,^^Jif you use 'upmath' package.^^J}%
      }
  \else
  \fi
\fi

\ifCUPmtlplainloaded \else
  \checkfont{msam10}
  \iffontfound
    \IfFileExists{amssymb.sty}
      {\typeout{^^JFound AMS Symbol fonts on the system, using the
                'amssymb' package.^^J}%
       \usepackage{amssymb}%

      }{}
  \fi
\fi

\ifCUPmtlplainloaded \else
  \IfFileExists{amsbsy.sty}
    {\typeout{^^JFound the 'amsbsy' package on the system, using it.^^J}%
     \usepackage{amsbsy}}
    {\providecommand\boldsymbol[1]{\mbox{\boldmath $##1$}}}
\fi

\providecommand\bnabla{\boldsymbol{\nabla}}
\providecommand\bcdot{\boldsymbol{\cdot}}

\newsavebox{\astrutbox}
\sbox{\astrutbox}{\rule[-5pt]{0pt}{20pt}}

\newcommand\shalf{\ensuremath{{\scriptstyle\frac{1}{2}}}}

\newcommand\thalf{\ensuremath{{\textstyle\frac{1}{2}}}}

\newcommand\ddint{\mathrm{d}}
\newcommand\icomp{\mathrm{i}}

\title[Segregation of a liquid mixture by a radially oscillating
bubble]%
{Segregation of a liquid mixture by a radially oscillating bubble}

\author[O. Louisnard, F. J. Gomez and R. Grossier]%
{O\ls L\ls I\ls V\ls I\ls E\ls R\ns %
L\ls O\ls U\ls I\ls S\ls N\ls A\ls R\ls D%
$^1$\ns, %
F\ls R\ls A\ls N\ls C\ls I\ls S\ls C\ls O\ns J.\ns %
G\ls O\ls M\ls E\ls Z%
$^2$%
\\\ns\and %
R\ls O\ls M\ls A\ls I\ls N\ns%
G\ls R\ls O\ls S\ls S\ls I\ls E\ls R%
$^1$}

\affiliation{$^1$Laboratoire de G\'enie des
  Proc\'ed\'es des Solides Divis\'es, Ecole des Mines d'Albi, 81013 ALBI
  Cedex 09, FRANCE\\[\affilskip]
  $^2$Laboratorio de Ultrasonidos, Dpto. de Fisica, Universidad de
  Santiago de Chile, Casilia 302, Santiago, CHILE}

\pubyear{}
\volume{}
\pagerange{}
\date{?? and in revised form ??}
\newcommand{\addedone}[1]{{#1}}%
\newcommand{\addedtwo}[1]{{#1}}%
\newcommand{\addedall}[1]{{#1}}%
\newcommand{\addedthree}[1]{{#1}}%

\newcommand\Pen{\mbox{\textit{Pe}}}  % Peclet number
\newcommand{\ds}{\displaystyle}
\newcommand{\vv}[1]{\boldsymbol{#1}}
\newcommand{\svit}{v}
\newcommand{\vit}{\vv{\svit}}
\newcommand{\ja}{\vv{j_A}}
\newcommand{\cgp}{{\mathcal R}}
\newcommand{\spec}[1]{\overline{#1}}

\newcommand{\vb}{\spec{v}_B}
\newcommand{\diml}[1]{\tilde{#1}}

\newcommand{\cad}{C}
\newcommand{\csm}{\cad^{sm}}

\newcommand{\cosc}{\cad^{osc}}
\newcommand{\coscinf}{\bar{\cad}^{osc}_1}
\newcommand{\cnum}{\cad^{num}}
\newcommand{\vp}{\dot{V}}
\newcommand{\vpp}{\ddot{V}}
\newcommand{\rp}{\dot{R}}
\newcommand{\rpp}{\ddot{R}}
\newcommand{\pe}{\Pen}
\newcommand{\undemi}{\frac{1}{2}}
\newcommand{\dsurd}[2]{\frac{\ds\partial #1}{\ds\partial #2}}
\newcommand{\sdsurd}[2]{\partial #1/\partial #2}
\newcommand{\ddesurd}[2]{\frac{\ds\partial^2 #1}{{\ds\partial #2}^2}}
\newcommand{\pinf}{p_{\infty}}
\newcommand{\cdim}{\omega_A}
\newcommand{\cdimz}{{\omega_A}_0}
\newcommand{\rz}{\diml{R}_0}
\newcommand{\ttiers}{\ensuremath{{\textstyle\frac{1}{3}}}}

\newcommand{\gsep}{G}
\newcommand{\ipemdemi}{\frac{1}{\pe^{1/2}}}
\newcommand{\itpemdemi}{\pe^{-1/2}}
\newcommand{\ipemun}{\frac{1}{\pe}}
\newcommand{\itpemun}{\pe^{-1}}
\newcommand{\pemdemi}{\pe^{1/2}}

\newcommand{\vqt}{V^{4/3}}
\newcommand{\vst}{V^{7/3}}

\newcommand{\taunl}{{\hat{\tau}}}
\newcommand{\pernl}{{\hat{T}}}
\newcommand{\abis}{A'}
\newcommand{\bbis}{B'}

\newcommand{\gfi}{F^i}
\newcommand{\gbi}{B^i}
\newcommand{\coscas}{{\bar{C}^{\text{osc}}}}

\newcommand{\czi}{\csm_{0,\infty}}
\newcommand{\cui}{\csm_{1,\infty}}
\newcommand{\cii}{\csm_{i,\infty}}

\newcommand{\rr}{\diml{R}}
\newcommand{\troisdemi}{\frac{3}{2}}
\newcommand{\ppg}{\diml{p}_g}

\newcommand{\oa}{\omega_A}

\newcommand{\rmin}{R_{\text{min}}}
\newcommand{\rpmin}{\rp_{\text{min}}}
\newcommand{\rpmax}{\rp_{\text{max}}}
\newcommand{\rppmax}{\rpp_{\text{max}}}

\newcommand{\hmin}{H_{\text{min}}}
\newcommand{\xmin}{x_{\text{min}}}
\newcommand{\taunlmin}{\taunl_{\text{min}}}
\newcommand{\dx}{\Delta x}
\newcommand{\gamundemi}{\Gamma\left({1}/{2}\right)}
\newcommand{\heav}{{\mathcal H}}
\newcommand{\fser}{F}
\newcommand{\fapp}{\fser_{\text{app}}}
\newcommand{\fmax}{\fser^{\text{max}}}
\newcommand{\fappmax}{\fapp^{\text{max}}}

\newcommand{\gser}{G}
\newcommand{\gapp}{\gser_{\text{app}}}
\newcommand{\gmin}{\gser^{\text{min}}}
\newcommand{\gappmin}{\gapp^{\text{min}}}

\newcommand\sgn{\mbox{sgn}} % 
\newcommand{\kgpermcube}{\nobreak\mbox{$\;$kg\,m$^{-3}$}}
\newcommand{\mpersec}{\nobreak\mbox{$\;$m\,s$^{-1}$}}

\newcommand{\mcarpersec}{\nobreak\mbox{$\;$m$^2$\,s$^{-1}$}}
\newcommand{\secpermcar}{\nobreak\mbox{$\;$s$^2$\,m$^{-2}$}}

\newcommand{\gpermol}{\nobreak\mbox{$\;$g\,mol$^{-1}$}}

\newcommand{\microns}{\nobreak\mbox{$\;\mu$m}}
\newcommand{\nm}{\nobreak\mbox{$\;$nm}}
\newcommand{\dalton}{\nobreak\mbox{$\;$Da}}

\newcommand{\viscunit}{\nobreak\mbox{$\;$kg\,m$^{-1}$\,s$^{-1}$}}

\newcommand{\presunitbis}{\nobreak\mbox{$\;$Pa} }
\newcommand{\tsurfunit}{\nobreak\mbox{$\;$N\,m$^{-1}$}}

\newcommand{\segratsm}{\Omega_m}
\newcommand{\segratosc}{\Delta\Omega_m}

\newcommand{\navogadro}{{\mathcal N}_a}
\newcommand{\sgnm}{\epsilon_m}
\begin{document}

\maketitle

\begin{abstract}
%  This concentration gradient at the bubble
%  wall yields an oscillatory concentration field in the whole liquid.
  A theoretical formulation is proposed for forced mass transport by
  pressure gradients in a liquid binary mixture around a spherical
  bubble undergoing volume oscillations in a sound field. Assuming the
  impermeability of the bubble wall to both species, diffusion driven
  by pressure gradients and classical Fick-diffusion must cancel at
  the bubble wall, so that an oscillatory concentration gradient
  arises in the vicinity of the bubble.  The P\'eclet number $\pe$ is
  generally high in typical situations and Fick diffusion cannot
  restore equilibrium immediately, so that an asymptotic average
  concentration profile may progressively build \addedthree{up} in the liquid over
  large times. Such a behavior is reminiscent of the so-called
  rectified diffusion problem, leading to slow growth of gas bubble
  oscillating in a sound field. A rigorous method formerly proposed by
  \cite{fyrillasszeri1} to solve the latter problem \addedthree{is} used in
  this paper to solve the present one. \addedthree{It is based on} splitting
  the problem \addedthree{into} a smooth part and an oscillatory part.  The smooth
  part is solved by a multiple scales method and yields the slowly
  varying average concentration field everywhere in the liquid.  The
  oscillatory part is obtained by matched asymptotic expansions in
  terms of the small parameter $\pe^{-1/2}$: the inner solution is
  required to satisfy the oscillatory balance between pressure
  diffusion and Fick diffusion at the bubble wall, while the outer
  solution is required to be zero. Matching both solutions yields a
  unique splitting of the problem. The final analytical solution,
  truncated to leading order, compares successfully to direct
  numerical simulation of the full convection--diffusion equation. The
  analytical expressions \addedthree{for} both smooth and oscillatory parts are
  calculated for various sets of bubble parameters: driving pressure,
  frequency and ambient radius.  The smooth problem always yields an
  average depletion of the heaviest species at the bubble wall, only
  noticeable for large molecules or nano-particles. For driving
  pressures sufficiently high to yield inertial oscillations of the
  bubble, the oscillatory problem predicts a periodic peak excess
  concentration of the heaviest species at the bubble wall at each
  collapse, lingering on several tens of time the characteristic
  duration of the bubble rebound. The two effects may compete for
  large molecules and practical implications of this segregation
  phenomenon are proposed for various processes involving acoustic
  cavitation.
 \end{abstract}

%%%%%%%%%%%%%%%%%%%%%%%%%%%%%%%%%%%%%%%%%%%%%%%%%%%%%%%%%%
\section{Introduction}
%%%%%%%%%%%%%%%%%%%%%%%%%%%%%%%%%%%%%%%%%%%%%%%%%%%%%%%%%%
Radially oscillating bubbles forced by a sound field are commonly
encountered in acoustic cavitation and sonoluminescence experiments
\cite[][]{sonochemandsonolumNAT}.  Generally, the liquid surrounding
such bubbles is not pure and involves various chemical species, which
may either participate in chemical reactions, or undergo phase
transitions, like crystallization processes. The kinetics of
such processes depends on the concentrations of the species and may
therefore be influenced by any variation of the
spatial \addedthree{homogeneity of the mixture}.

Pressure gradients may be a possible source of mixture segregation.
Following diffusion theory \cite[][]{hirshfelder,bird}, when a mixture
of two species is \addedthree{subjected} to a pressure gradient, the
lightest one is pushed toward low pressure regions. This forced
diffusion process, known as pressure diffusion, \addedthree{generally
  remains} weak unless the liquid is submitted to high pressure
gradients, as in ultracentrifuge applications where it is used
profitably to separate large molecular weight species from a solvent
\cite[][]{archibald38}. Pressure diffusion is also responsible for
\addedthree{gas} stratification in a \addedthree{quiescent}
atmosphere, or solute-solvent segregation in long sedimentation
columns \cite[][]{mullinleci69,larsongarside86}. Besides, the effect
of pressure diffusion, along with thermal diffusion, on the
segregation of a gas mixture inside a radially oscillating bubble has
been investigated by \cite{storeyszerimixture} in the context of
sonoluminescence.

When a bubble is driven in radial motion by a high amplitude
oscillating pressure field, pressure gradients arise in the liquid, as
a result of the bubble wall acceleration. Inertial cavitation \addedthree{is}
a situation where the bubble suffers an explosive expansion followed
by a violent collapse. In \addedthree{this} case, the pressure gradient reaches a
very high value near the end of the collapse, owing to the strong gas
compression in the bubble.  The segregation of two species by pressure
diffusion in the neighborhood of a cavitation bubble may therefore
\addedthree{notably influence} the liquid homogeneity.

The similar problem of mass transport of a gas dissolved in a liquid
around a bubble undergoing volume oscillations has been studied
extensively
\cite[][]{ellerflynn,hsiehplesset,fyrillasszeri1,fyrillasszeri2}: the
variations of the gas concentration at the bubble wall, driven by the
bubble oscillations, yield a non-zero average gas flux toward the
bubble, reversing its natural dissolution process, a phenomenon known
as rectified diffusion. In this case, the dissolved gas flux in the
liquid arises as a consequence of the asymmetry in the behavior of
the two components at the bubble wall: the gas can cross the
\addedthree{interface}, the liquid cannot.  In the present problem,
assuming a binary mixture of two non-volatile and non-surface-active
fluids, the bubble \addedthree{interface} acts as a barrier for the
two species, which would prohibit any relative flux.  However, if
pressure diffusion is taken into account, a new asymmetry between the
two components arises, owing to their different densities, and the
pressure gradient near the bubble wall separates the two species.
Since the net flux across the bubble wall of any of the two species
must be zero, a non-zero Fick diffusion flux must \addedthree{exactly
  balance} the pressure diffusion flux. Thus, a concentration gradient
should appear near the bubble wall and by continuity in the whole
liquid. Since the pressure gradient reverses as the bubble oscillates,
it is clear that the concentration of each species is an oscillatory
quantity, but the question also arises of a possible average effect,
building over several periods, as observed for rectified diffusion.

In the present paper, an approximate analytical expression of the
concentration field in the liquid is sought in order to be able to
draw some conclusions for a given mixture and given bubble parameters,
namely the amplitude of the driving pressure, its frequency and the
ambient radius of the bubble. Since the problem has some common
characteristics with rectified diffusion, the solution method proposed
by \cite{fyrillasszeri1} will be used. The concentration field is cut
in two \addedthree{parts}: the oscillating field is required to
fulfill the complicated oscillatory part of the boundary condition at
the bubble wall and is non-zero only in a thin boundary layer near the
bubble; the smooth field satisfies the remaining part of the boundary
condition and is uniformly valid everywhere in the liquid. No specific
assumption is made \addedthree{concerning bubble dynamics}, apart from
its periodic motion, so that the solution obtained is immediately
applicable once the bubble radius is known as a function of time.

The paper is organized as follows: section 2 presents the main
convection--diffusion equation along with boundary conditions and the
splitting of the problem in two parts. In section 3, the oscillatory
problem is solved and the splitting is determined unambiguously. The
smooth problem is solved in section 4. In section 5, the analytical
results are first validated by comparing them to a full numerical
solution of the partial differential equation. Then, the influence of
the bubble parameters on the magnitude of the segregation effect is
investigated. In section 6, the model is finally applied to typical
mixtures of water with either small or large molecules and the results
are discussed.
  
%%%%%%%%%%%%%%%%%%%%%%%%%%%%%%%%%%%%%%%%%%%%%%%%%%%%%%%%%%
\section{Formulation}
%%%%%%%%%%%%%%%%%%%%%%%%%%%%%%%%%%%%%%%%%%%%%%%%%%%%%%%%%%

%=========================================================
\subsection{Bubble motion and liquid fields}
%========================================================= 
We will consider a single bubble oscillating in a liquid mixture of
infinite extent, forced by a oscillating pressure field far from the
bubble $\pinf(t) = p_0(1-P\cos \omega t)$, where $\omega$ is the
angular frequency, $P$ the dimensionless forcing pressure and $p_0$
the hydrostatic pressure. The motion of such a bubble has been widely
described in the literature since the early work of Lord Rayleigh and
several refinements of the basic model can be found, including thermal
behavior of the gas, liquid compressibility effects and liquid
evaporation at the bubble wall \cite[see][for a recent
review]{prospernato,brenner2002}.

The model presented \addedthree{here} is in itself independent of a specific
choice \addedthree{of} a bubble dynamics model, and we defer the choice of the
differential equation governing the radial motion to section
\ref{secresults}.  However, the mass transport equation used \addedthree{in}
this work requires analytical expressions of the velocity and pressure
fields in the liquid, and for the sake of simplicity, we will assume
an iso-volume motion of the liquid. Besides, the potential character
of the flow is ensured by the spherical symmetry, and the potential
$\diml{\phi}$ and velocity fields $\diml{\svit}$ can be easily
obtained from mass conservation:
 \begin{eqnarray}
  \label{potentiel}
  \diml{\phi}(r,t) &=& -\frac{1}{4\pi r}\frac{d\diml{V}}{dt},\\
  \label{vitesse}
  \diml{\vit}(r,t) &=& \frac{\vv{r}}{4\pi r^3} \frac{d\diml{V}}{dt},
\end{eqnarray}
where $\diml{V}$ is the time-dependent bubble volume, and $r$ the
distance from the center of the bubble. Then, using the unsteady
Bernoulli law for potential flows between a point in the liquid of
radial coordinate $r$ and a point infinitely far from the bubble, the
pressure field in the liquid is
\begin{equation}
  \label{defpresdim}
  \diml{p}(r,t) = \pinf(t) + 
  \rho\left[
    \frac {1} {4\pi r} \frac{d^2\diml{V}}{dt^2}
    -\frac{1}{32\pi^2 r^4}\left(\frac{d\diml{V}}{dt} \right)^2
  \right].
\end{equation}
The validity of the iso-volume assumption \addedthree{is questionable}
for strong motion of the bubble, involving wall velocities near or
greater than the speed of sound in the liquid.  Accounting for liquid
compressibility results in corrections of the order of the Mach number
in both the velocity and pressure field, and therefore also in the
bubble dynamics equation \cite[][]{prosperlezzi1}. \addedtwo{The main
  physical consequence of liquid compressibility is the formation of
  shock-waves at the end of the bubble collapse
  \cite[][]{hicklingplesset}. Clearly, since shock-waves are in
  essence strong pressure gradients, neglecting their effect on
  pressure diffusion may appear as a rough approximation. However,
  taking them into account would require cumbersome expressions of the
  velocity and pressure fields \cite[see for example the second-order
  expressions of ][obtained by the PLK strained-coordinates
  method]{tomitashima,fujikawa}. This constitutes a technical problem
  especially for the velocity field: as will be seen below, the
  convective term in the transport equation can be easily suppressed
  by a convenient change of variable in the incompressible case. There
  is no evidence that a similar change of variable could be found in
  the case of a compressible velocity field, which would make the
  problem untractable. We therefore chose to sacrifice the
  compressibility hypothesis in order to draw a general picture of the
  pressure diffusion effect. We will however make an exception and
  keep the compressibility-induced correction terms in the bubble
  equation, in order to obtain a more realistic model for the bubble
  dynamics.  Moreover, since the shock-waves issue is of practical
  interest, additional qualitative comments will be proposed in
  section 6.}

Finally, since the liquid considered here is a mixture whose spatial
homogeneity is investigated, the average mixture density may not be
constant. \addedtwo{It may be reasonably assumed that the occurrence
  of such inhomogeneities does neither significantly affect the liquid
  fields, nor the bubble motion}.
% , so that here and in what follows, $\rho$ will represent the
% mixture density at rest.% Using equation (\ref{defpresdim}) in the expression of the
% pressure-jump across the bubble-wall yields the well-known
% Rayleigh-Plesset equation:
% %
% \begin{equation}
%   \label{rayleigh}
%   \diml{R}\frac{d^2\diml{R}}{dt^2} 
%   + \frac{3}{2}\left(\frac{d\diml{R}}{dt} \right)^2
%   =\frac{1}{\rho}\left[p_g(t)
%   -\frac{2\sigma}{\diml{R}} -\frac{4\mu}{\diml{R}}\frac{d\diml{R}}{dt} 
% -{\pinf}(t)\right].
% \end{equation}
% %
% Taking into account the compressibility of the liquid would yield more
% complicated fields and bubble equations. Such corrections have been
% widely discussed in the literature \cite[see][and references
% herein]{prosperlezzi1,brenner2002}. The results we develop in the
% present work, although irrespective of a specific choice for bubble
% dynamics, rely on the incompressible velocity and pressure fields
% (\ref{vitesse}) ans (\ref{defpresdim}).  For simplicity we will carry
% on with this assumption.  In equation (\ref{rayleigh}), 
%
% where $\pinf(t) = p_0(1-P\cos \omega t)$ is the oscillating sound
% pressure in the liquid, with $\omega$ angular frequency, $P$
% dimensionless acoustic pressure and $p_0$ hydrostatic pressure; $\mu$
% is the viscosity of the liquid; $\sigma$ is the surface tension.
 
%=========================================================
\subsection{Mass transport}
%=========================================================
The mass conservation of a species A in a binary mixture is expressed as
\begin{equation}
  \label{conserv}
  \dsurd{\rho_A}{t} + \bnabla \bcdot (\rho_A\diml{\vit}) = -\nabla \bcdot {\ja},
\end{equation}
where $\rho_A$ is the local density of species A, $\diml{\vit}$ is
the mass-averaged mixture velocity. The mass diffusion flux $\ja$
is, taking into account pressure diffusion
\cite[][]{hirshfelder,bird}
\begin{equation}
  \label{defja}
  \ja = - D\left[\rho\bnabla{\omega_A} + 
    \frac{M_A M_B}{M\cgp T}\rho\omega_A
    \left(\frac{\bar{V}_A}{M_A}-\frac{1}{\rho} \right)
    \bnabla{\diml{p}}
  \right],
\end{equation}
where $\omega_A=\rho_A/\rho$ is the mass fraction of species A, $M_A$,
$M_B$ and $M$ are the respective molar weight of species A, B and of
the mixture, $\rho$ is the mean density of the mixture, $\bar{V}_A$
the specific molal volume of species A, and $\cgp$ the universal gas
constant.  The first term in equation (\ref{defja}) represents the
Fick diffusion flux driven by a concentration gradient, and the second
is the pressure diffusion flux, driven by the local pressure gradient.

Using the mixture mass-conservation equation
$\sdsurd{\rho}{t}+\bnabla\bcdot{(\rho\diml{\vit})} = 0$, and inserting
the flux expression (\ref{defja}) in equation (\ref{conserv}), we get:
\begin{equation}
  \label{conservomgen}
  \rho\left(\dsurd{\omega_A}{t} + \diml{\vit}\bcdot\bnabla\omega_A \right) = 
  D \bnabla\bcdot\left\{
    \rho\left[\bnabla\omega_A + 
      \frac{M_A M_B}{M\cgp T}\omega_A\left(\frac{\bar{V}_A}{M_A}-\frac{1}{\rho} \right)
      \bnabla{\diml{p}}\right]
  \right\}.
\end{equation}
Although \addedthree{it is tempting} to simplify both sides of equation
(\ref{conservomgen}) by $\rho$, the latter quantity is not constant
since it depends on the local mass fraction $\omega_A$, which is space
and time dependent.  
% This simplification is generally done without
% further justification in ultracentrifugation problems
% \cite[][]{archibald38}. 
The same problem arises with the appearance of the mean molar weight
$M$ and again the density $\rho$ in the pressure diffusion term of
equation (\ref{conservomgen}), which depends on the local composition
of the mixture. This has the strong consequence that rigorously,
equation (\ref{conservomgen}) is non-linear. In view of the method we
plan to use for the resolution of the problem, a linearization of the
problem is necessary, paying the price of some additional assumptions.
It is shown in appendix \ref{annlinear} that in the limit of a dilute
mixture ($\omega_A\ll 1$), the mixture density $\rho$ is approximately
constant and equal to $M_B/\bar{V}_B$ and equation
(\ref{conservomgen}) may be simplified as
\begin{equation}
  \label{conservomdilue}
  \dsurd{\cdim}{t} + \diml{\vit}\bcdot\bnabla\cdim  = 
  D\bnabla\bcdot\left[\bnabla{\cdim} + 
    \frac{M_A}{\cgp T} \cdim
    \left(\frac{\bar{V}_A}{M_A}-\frac{\bar{V}_B}{M_B} \right)
    \bnabla{\diml{p}}
  \right],
\end{equation}
or, in spherical coordinates:
\begin{equation}
  \label{conservspher}
  \dsurd{\cdim}{t} + \diml{\svit}(r,t)\dsurd{\cdim}{r}  = 
  \frac{D}{r^2}\dsurd{}{r}
  \left[ r^2 
    \left( \dsurd{\cdim}{r} 
      + \diml{\beta} \cdim\dsurd{\diml{p}}{r}(r,t) \right)
  \right],
\end{equation}
where $\diml{v}(r,t)$ and $\diml{p}(r,t)$ are given by
(\ref{vitesse}),(\ref{defpresdim}), and $\diml{\beta}$ is:
\begin{equation}
  \label{defbetadim}
  \diml{\beta} = \frac{M_A}{\cgp T}
    \left(\frac{\bar{V}_A}{M_A}-\frac{\bar{V}_B}{M_B} \right).
\end{equation}
This conservation equation should be completed with appropriate
boundary conditions at the bubble wall and infinitely far from the
bubble. Both \addedthree{components} are assumed non-volatile and
therefore cannot cross the bubble wall.  Thus, the diffusive flux
$\ja$ should be zero at the bubble wall:
\begin{equation}
  \label{boundaryR}
  \dsurd{\cdim}{r}(r=\diml{R}(t),t) + \diml{\beta}
  \cdim(r=\diml{R}(t),t)\dsurd{\diml{p}}{r}(r=\diml{R}(t),t)
  = 0.
\end{equation}
We emphasise that the expression of the diffusive flux at the bubble
wall must include the pressure diffusion term, consistently with the
transport equation (\ref{conservspher}). It is interesting to note
that a similar boundary condition is used in centrifuge equations
\cite[][]{archibald38}, to express the impermeability of the
sample-tube extremities to any species.  \addedone{The non-volatility
  of the species may appear as a drastic limitation. However, relaxing
  this hypothesis would have the disadvantage to couple the diffusion
  problem in the liquid with the diffusion problem of vapor through
  the uncondensable gas filling the bubble. Moreover, the problem
  would require a liquid-vapor equilibrium condition at the bubble
  wall, which may take a complex form in the case of mixtures.
  Finally, several related issues such as evaporation--condensation
  kinetics, or chemical reactions \cite[][]{storeyszeri2000} may
  further complicate the problem. We therefore leave aside these
  refinements for now, and concentrate on the effects of pressure
  diffusion alone.}

Far from the bubble, the concentration field remains undisturbed by
the bubble oscillations, so that
\begin{equation}
  \label{boundaryinf}
  \cdim (r\rightarrow\infty,t) = \cdimz, 
\end{equation}
and finally, the liquid mixture is initially assumed homogeneous in
space:
\begin{equation}
  \label{initial}
   \cdim (r,t=0) = \cdimz.
\end{equation}
% 
%=========================================================
\subsection{Non-dimensionalization}
%=========================================================
The equations of the problems are non-dimensionalized as follows: the
natural length scale is the ambient bubble radius $\rz$, the time scale
is $\omega^{-1}$, the inverse of the driving frequency. The pressure
scale is set as $\thalf\rho_0 \rz^2\omega^2$, which is the dynamic
pressure of the liquid displaced by the bubble. We therefore set
\[
  r = \rz \xi, 
  \quad t =  \tau/\omega, 
  \quad \diml{p} = \frac{1}{2}\rho \rz^2\omega^2 p.
\]
The bubble instantaneous radius and volume are non-dimensionalized by
their ambient values:
\[
  \label{vrad}
  \diml{R} = \rz R, \quad \diml{V} = \frac{4}{3}\pi \rz^3 V.
\]
In the new variables, the dimensionless velocity and pressure field in
the liquid are
\begin{subequations}
  \begin{equation}
    \label{vadim}
    v(\xi,\tau) = \frac{\vp}{3\xi},
  \end{equation}
  \begin{equation}
    \label{presadim}
    p(\xi,\tau) = \frac{2}{3}\frac{\vpp}{\xi}-\frac{1}{9}\frac{\vp^2}{\xi^4},
  \end{equation}
\end{subequations}
where \addedthree{here,} and in what follows, over-dots
\addedthree{denote} time-derivatives with respect to the dimensionless
time-variable $\tau$.  The concentration of species A is
non-dimensionalized by
\begin{equation}
  \label{cadim}
  \cad = \frac{\cdim}{\cdimz} - 1.
\end{equation}
The quantity $\cad$ represents the segregation level: a positive value
of $\cad$ expresses a local excess of species A above the concentration
at rest. The mass transport-equation becomes
\begin{equation}
  \label{conservomad}
  \dsurd{\cad}{\tau} + \left(\frac{\vp}{3\xi^2} \right) \dsurd{\cad}{\xi}
  = \frac{1}{\pe}\frac{1}{\xi^2}\dsurd{}{\xi}\left\{
  \xi^2\left[\dsurd{\cad}{\xi} + \beta (\cad+1)\dsurd{p}{\xi} \right]
\right\},
\end{equation}
where $\pe=\rz^2\omega/D$ is the P\'eclet number, and the dimensionless
number $\beta$ is 
\begin{equation}
  \label{defbetad}
  \beta = \diml{\beta}\undemi\rho \rz^2\omega^2.  
\end{equation}
For later use, we separate the respective contributions of the mixture
and the bubble to the dimensionless parameter $\beta$, and write
\begin{equation}
  \label{defbetadbis}
  \beta = \beta_m \rz^2\omega^2, 
\end{equation}
where
\begin{equation}
  \label{defbetam}
  \beta_m =
  \undemi\diml{\beta}\rho
  =   \undemi \frac{M_A}{\cgp T}
  \left(\frac{\bar{V}_A}{M_A}\frac{M_B}{\bar{V}_B} - 1 \right),
\end{equation}
depends only on the mixture considered.

The boundary and initial conditions \addedthree{become}, in dimensionless
variables
\begin{subeqnarray}
  \label{boundinad}
  \slabel{boundaryRad}
    & &\dsurd{\cad}{\xi}(\xi=R(\tau),\tau) + 
    \beta  \left[\cad(\xi=R(\tau),\tau)+1 \right]
    \dsurd{p}{\xi}(\xi = R(\tau),\tau)  = 0, \\
    \slabel{boundaryinfad}
    & & \cad(\xi\rightarrow\infty,\tau) = 0, \\
    \slabel{initialad}
    & & \cad(\xi,\tau=0) =  0.
\end{subeqnarray}
The intrinsic difficulties in the resolution of the above governing
equations are similar to \addedthree{those} encountered in the
rectified diffusion problem
\cite[][]{hsiehplesset,ellerflynn,fyrillasszeri1,fyrillasszeri2}:
\addedthree{on} one hand, the boundary condition (\ref{boundaryRad})
at the bubble surface is applied at a moving boundary and is
furthermore unsteady.  \addedthree{On} the other hand, the velocity field is
inhomogeneous and also unsteady.  The solution to overcome the
difficulty of the moving boundary and the oscillating velocity field
is to define a Lagrangian radial coordinate, as first suggested by
\cite{plessetzwick}, by $\sigma=\ttiers(\xi^3-V(\tau))$, which
represents physically the dimensionless volume between the bubble wall
and the point of interest in the liquid. A specific liquid particle
moves with a constant $\sigma$ (under the incompressibility
hypothesis) and an observer moving with such a particle would only see
the diffusive transport of species. This may be readily seen by
expressing equation (\ref{conservomad}) in the Lagrangian coordinates
$(\sigma,\tau)$:
\begin{equation}
  \label{conservfin}
  \dsurd{\cad}{\tau}  = \frac{1}{\pe}\dsurd{}{\sigma}
  \left[
    A(\sigma,\tau) \dsurd{\cad}{\sigma} + 
     \beta B(\sigma,\tau) (\cad+1)\right],
\end{equation}
where
\begin{subequations}
  \label{defAB}
  \begin{equation}
    \label{defA}
    A(\sigma,\tau) = (3\sigma+V)^{4/3},     
  \end{equation}
  \begin{equation}
    \label{defB}
    B(\sigma,\tau) = -\frac{2}{3}\vpp+\frac{4}{9}\frac{\vp^2}{(3\sigma+V)}.
  \end{equation}
\end{subequations}
The boundary and initial conditions (\ref{boundinad}) become
\begin{subeqnarray}
  \label{boundinlag}
  \slabel{boundaryRag}
  & &\dsurd{\cad}{\sigma}(0,\tau) +
  \beta \frac{B(0,\tau)}{A(0,\tau)} (\cad(0,\tau)+1) = 0, \\
  \slabel{boundaryinflag}
  & &\cad(\sigma\rightarrow\infty,\tau) = 0, \\
  \slabel{initiallag}
  & &\cad(\sigma,\tau=0) =  0.
\end{subeqnarray}
It can be readily seen that the convective term of equation
(\ref{conservomad}) has indeed disappeared and that the boundary
condition at the bubble wall is now applied at a fixed point, thanks
to the change of variable.

The problem defined by equations (\ref{conservfin})-(\ref{boundinlag})
shares some resemblance with the rectified diffusion problem and thus,
the splitting-method proposed by
\cite{fyrillasszeri1,fyrillasszeri2,fyrillasszerisurfac} can be
profitably used here.  For self-completeness, we will recall here the
main lines of its underlying physical basis. For rectified diffusion,
the oscillatory gas pressure in the bubble drives the gas concentration
in the neighbouring liquid in oscillation, thus producing a periodic
inversion of the concentration gradient. This rapidly oscillating
gradient is counteracted by molecular diffusion, but owing to the
large value of the P\'eclet number, the equilibrium cannot be restored
immediately.  This delay produces a long-term average diffusion
effect, on a timescale larger than the oscillation period by a factor
of the order of $\pe$. Therefore the concentration field varies on two
time-scales.

The present problem shares this property with rectified diffusion, but
here, the source of \addedthree{the} concentration gradient is the
segregation of species by pressure diffusion, both in the liquid bulk
and at the bubble wall (see equations (\ref{conservfin}) and
(\ref{boundaryRag}) respectively). Moreover, it can be
\addedthree{easily seen} by looking at equation (\ref{presadim}) or
(\ref{defB}) that pressure diffusion is not symmetric over one
oscillation period, owing to the $\vp^2$ term, representing the
convective acceleration of the fluid in spherical symmetry. Because of
this term, pressure increases when traveling away from the bubble,
which may be understood \addedthree{from the} Bernoulli law: because
of spherical symmetry, velocity decreases with the distance to the
bubble and this should be balanced by a pressure increase.

For both rectified diffusion and the present problem, the existence of
two time scales justifies a multiple-scales method, but there remains
a technical difficulty in the unsteady character of the boundary
conditions at the bubble \addedthree{wall. If} the multiple-scales method were to
be applied directly to the set of equations
(\ref{conservfin}),(\ref{boundinlag}), one would be met with an
impossibility for the solution at leading order to fulfill the
oscillating boundary condition (\ref{boundaryRag}). This difficulty is
overcome by splitting the concentration field in two parts: an
oscillatory \addedthree{part}, which satisfies the oscillating part of the boundary
condition, and a smooth \addedthree{part}, to which the remaining part of the
boundary condition should be ascribed. The oscillating part represents
physically the perturbation of the concentration field imposed by the
bubble wall forcing term, and is designed to differ from zero only in
a boundary layer of thickness $\pe^{-1/2}$. The smooth part extends in
the whole liquid and varies on both the oscillation timescale
$1/\omega$ and a slow timescale of order $\pe/\omega$.

%=========================================================
\subsection{Splitting of the problem}
%=========================================================
We set the concentration field as $\cad(\sigma,\tau) =
\cosc(\sigma,\tau) + \csm(\sigma,\tau)$, where $\cosc$ is the
oscillatory part, and $\csm$ the smooth part. We then split the
governing equations \addedthree{into} an oscillatory part and a smooth
\addedthree{part}. The oscillatory problem is defined by
\begin{subeqnarray}
  \label{osc}
  \slabel{eqosc}
  & &\dsurd{\cosc}{\tau}  = \frac{1}{\pe}\dsurd{}{\sigma}
  \left[
    A(\sigma,\tau) \dsurd{\cosc}{\sigma} + 
    \beta B(\sigma,\tau) \cosc  \right], \\
  \slabel{closc}
  & &\dsurd{\cosc}{\sigma}(0,\tau) + 
  \beta H(\tau) \left[\cosc(0,\tau) + 1\right] = - 
  \gsep - \beta H(\tau)\csm(0,\tau) ,
\end{subeqnarray}
%                                %
and the smooth problem is
\begin{subeqnarray}
  \label{smooth}
  \slabel{eqsmooth}
  & &\dsurd{\csm}{\tau}  = \frac{1}{\pe}\dsurd{}{\sigma}
  \left[
    A(\sigma,\tau) \dsurd{\csm}{\sigma} + 
    \beta B(\sigma,\tau) \left(\csm + 1 \right)\right],\\
  \slabel{clsmooth}
  & &\dsurd{\csm}{\sigma}(0,\tau) = \gsep.
\end{subeqnarray}
The constant $\gsep$ is introduced to add a degree of freedom in the
separation process and will be determined unambiguously by using a
splitting condition, to be defined in the next section.  In the
boundary condition (\ref{closc}), the function $H(\tau)$ is defined by
\begin{equation}
  \label{defH}
  H(\tau) = \frac{B(0,\tau)}{A(0,\tau)},
\end{equation}
and represents the dimensionless pressure gradient at the bubble wall.

Finally, both oscillatory and smooth fields are required to fulfill
the boundary condition far from the bubble (\ref{boundaryinflag}) and
the initial condition (\ref{initiallag}), so that
 \begin{eqnarray}
   \label{clinfsm}
   \csm(\sigma\rightarrow\infty, \tau) &= \csm(\sigma,0) = 0, \\
   \label{clinism}
   \cosc(\sigma\rightarrow\infty, \tau) &= \cosc(\sigma,0) = 0.
 \end{eqnarray}
 
%%%%%%%%%%%%%%%%%%%%%%%%%%%%%%%%%%%%%%%%%%%%%%%%%%%%%%%%%%
\section{The oscillatory problem}
%%%%%%%%%%%%%%%%%%%%%%%%%%%%%%%%%%%%%%%%%%%%%%%%%%%%%%%%%%
% %%%%%%%%%%%%%%%%%%%%% A RECASER ? %%%%%%%%%%%%%%%%%%%%%%%%%%%%%%%%
% The splitting should be chosen so that $\cosc$, the oscillatory part
% of the approximation vanishes outside a thin layer near the bubble
% wall. This requisite allows the election of the precise splitting
% among an infinite number of possibilities. 
% %%%%%%%%%%%%%%%%%%%%% A RECASER ? %%%%%%%%%%%%%%%%%%%%%%%%%%%%%%%%
Following \cite{fyrillasszeri2}, we use a matched asymptotic expansion
to solve the oscillatory problem: the inner solution must fulfill the
bubble wall boundary condition (\ref{closc}) while the outer solution
is required to be identically zero.  
% As shown below and in appendix \ref{annosc}, this determines the
% splitting constant $\gsep$ in equations (\ref{closc}) and
% (\ref{clsmooth}).
To determine the inner approximation of the oscillatory solution, we
define a re-scaled Lagrangian space-coordinate by $s =
\pe^{1/2}\sigma$. Furthermore, we use the nonlinear time $\taunl$
first suggested by \cite{plessetzwick}, which arises from the
spherical symmetry of the problem:
\begin{equation}
  \label{deftaunl}
  \taunl = \int_0^\tau R^4(\tau') \:\ddint \tau',
\end{equation}
and for further use, we also define the nonlinear period
\begin{equation}
  \label{defpernl}
  \pernl = \int_0^{2\pi} R^4(\tau') \:\ddint \tau'.
\end{equation}
Taking $\taunl$ as the new time-variable, equation (\ref{eqosc}) becomes
\begin{equation}
  \label{eqoscvars}
    \dsurd{\cosc}{\taunl} = \dsurd{}{s} \left[
    \abis(s,\taunl;\pe) \dsurd{\cosc}{s} +
    \ipemdemi\beta \bbis(s,\taunl;\pe) 
     \cosc\right],
\end{equation}
where
\begin{subeqnarray}
  \label{defabbis}
  \slabel{defabis}
  \abis(s,\taunl;\pe) &=& R^{-4}A(\itpemdemi s,\tau)=\left(1+\frac{1}{\pemdemi}\frac{3s}{V} \right)^{4/3}, \\
  \slabel{defbbis} \bbis(s,\taunl;\pe) &=& R^{-4}B(\itpemdemi s,\tau)=
  -\frac{2}{3}\frac{\vpp}{\vqt}+\frac{4}{9}\frac{\vp^2}{\vst}
  \frac{1}{\ds 1+\ipemdemi \frac{3s}{V}},
\end{subeqnarray}
and the bubble wall condition reads
\begin{equation}
  \label{closcvars}
    \pemdemi \dsurd{\cosc}{s}(0,\taunl) + 
   \beta H(\taunl) \left[\cosc(0,\taunl) + 1\right] = 
   - \gsep - \beta H(\taunl)\csm(0,\taunl).
\end{equation}
The outer limit of the inner approximation should match the outer
approximation which is identically zero, so that $\cosc(s,\taunl)$
should satisfy
\begin{equation}
  \label{outerzero}
  \lim_{s\rightarrow\infty}\cosc(s,\taunl) = 0.
\end{equation}
We now assume an asymptotic expansion for $\cosc$ in the $\pe^{-1/2}$
parameter:
\begin{equation}
  \label{devcosc}
  \cosc(s,\taunl) = \cosc_0(s,\taunl) + \ipemdemi \cosc_1 (s,\taunl)
  + \frac{1}{\pe} \cosc_2(s,\taunl) \dots,
\end{equation}
and we also expand functions $\abis$ and $\bbis$ given by equations
(\ref{defabbis}), as well as the separation constant $\gsep$ appearing
in equations (\ref{closc}) and(\ref{smooth}):
\begin{eqnarray*}
  \ds  \abis(s,\taunl) &=& 1 + \itpemdemi \abis_1 (s,\taunl)
  + \itpemun \abis_2 (s,\taunl) \dots, \\
  \ds  \bbis(s,\taunl) &=& \bbis_0 (\taunl) + \itpemdemi \bbis_1 (s,\taunl)
  + \itpemun \bbis_2 (s,\taunl) \dots, \\
  \ds  \gsep &=& \gsep_0 + \itpemdemi \gsep_1 + \itpemun \gsep_2 \dots.
%  \ds  \fsep &=& \fsep_0 + \itpemdemi \fsep_1 + \itpemun \fsep_2 \dots.
\end{eqnarray*}
A hierarchy of inhomogeneous diffusion problems is obtained, all
sharing the same form. The general solution of such problems is
detailed in appendix~\ref{annosc}, which also yields a
splitting-condition necessary to ensure the matching equation
(\ref{outerzero}). It is interesting to note that in the present
\addedthree{case}, the oscillatory problem at each order has a Neumann
boundary condition, whereas the oscillatory problems in the analysis
of surfactants-enhanced rectified diffusion by \cite{fyrillasszeri2}
involve a Dirichlet boundary condition.  The difference arises from
the presence of the P\'eclet number in the boundary condition in the
latter problem \cite[see equation (2.4) in][]{fyrillasszeri2}, while
the boundary condition (\ref{boundaryRag}) in the present problem is
P\'eclet \addedthree{independent. This} is because both pressure and Fick diffusion
terms are proportional to the diffusion coefficient, which thus
cancels out in the null total flux condition (\ref{boundaryR}) at the
bubble wall.

For further use, it is useful to note that the $H$ function given by
(\ref{defH}) can also be expressed in the following forms   
\begin{equation}
  \label{relABH}
  H(\taunl) = \bbis_0(\taunl) =
  -\frac{2}{3}\frac{\vpp}{\vqt}+\frac{4}{9}\frac{\vp^2}{\vst}
  = -2\frac{\rpp}{R^2}.
\end{equation}
We now turn to solve the oscillatory problems at each order. 
%=========================================================
\subsection{Zeroth-order}
%=========================================================
The oscillatory problem at order 0 is
\begin{subequations}
  \label{pbmosc0}
  \begin{equation}
    \ds\dsurd{\cosc_0}{\taunl} = \ddesurd{\cosc_0}{s} 
    \slabel{eqosc0},
  \end{equation}
  \begin{equation}
    \label{clbulosc0}
  \ds\dsurd{\cosc_0}{s}(0,\taunl) = 0,
  \end{equation}
  \begin{equation}
    \label{clinfosc0}
    \cosc_0(s\rightarrow\infty) = 0.
  \end{equation}
\end{subequations}
The solution is clearly the null one. This can be easily understood on
a physical basis \addedthree{as} pressure diffusion does not act to this order,
neither in the liquid bulk, nor at the bubble wall, as can be seen in
(\ref{eqosc0}). Therefore the liquid mixture is only submitted to
classical molecular diffusion. Only a non-homogeneous boundary
condition could produce a concentration gradient which is not the
case, since to this order, the bubble wall condition only imposes a
zero concentration gradient. This is why, contrarily to rectified
diffusion problems \cite[][]{fyrillasszeri1,fyrillasszeri2}, the
zeroth-order oscillatory solution is zero in the present problem.

%=========================================================
\subsection{First-order}
%=========================================================
At order 1, using the nullity of $\cosc_0(0,\taunl)$, we obtain:
\begin{subequations}
  \slabel{pbmosc1}
  \begin{equation}
    \label{eqosc1}
    \ds \dsurd{\cosc_1}{\taunl} = \ddesurd{\cosc_1}{s},
  \end{equation}
  \begin{equation}
    \label{clbulosc1}
    \ds \dsurd{\cosc_1}{s}(0,\taunl) + \beta H(\taunl)= 
    -\gsep_0 -  \beta H(\taunl)\csm_0(0,\taunl),
  \end{equation}
  \begin{equation}
    \label{clinfosc1}
    \cosc_1(s\rightarrow\infty) = 0.
  \end{equation}
\end{subequations}
The splitting-condition (\ref{splitcond}) obtained in appendix
\ref{annosc} yields the separation constant $\gsep_0$:
\begin{equation}
  \label{defgsep0}
  \gsep_0 = - \beta
    \left< 
      H(\taunl)\left[
        \csm_0(0,\tau) +1
      \right] 
    \right>_{\taunl}.
\end{equation}
The part of the boundary condition ascribed to $\cosc_1$ is therefore
\begin{equation*}
  \dsurd{\cosc_1}{s}(0,\taunl) = \beta\left[
    \left< H(\taunl)\right>_{\taunl} - H(\taunl) 
  \right]
  \left[
    \csm_0(0) + 1
  \right],
\end{equation*}
where we have used the result, to be demonstrated in section
\ref{secsmooth}, that $\csm_0$ is independent of the fast
time-variable $\taunl$. The asymptotic solution $\coscinf(s,\taunl)$
of equations (\ref{pbmosc1}\,\textit{a--c}) can be obtained from
appendix \ref{annosc}: expanding $H(\taunl)$ as a Fourier series,

\begin{equation}
  \label{devfourierH}
  H(\taunl) =     \left< H(\taunl)\right>_{\taunl}
  +   \sum_{%
    \substack{
      m = -\infty \\
      m \neq 0}}%
  ^{m=+\infty}
\addedall{h_m \exp\left( 2\icomp m\pi \frac{\taunl}{\pernl} \right)},
\end{equation}
and using equation (\ref{coscnofi}), the oscillatory concentration field is
\begin{multline}
  \label{coscresult}
  \coscinf(s,\taunl) = \beta \left[ \csm_0(0) + 1 \right]
  \left(\frac{\pernl}{2\pi} \right)^{\shalf} \\
  \times  \sum_{%
    \substack{
      m = -\infty \\
      m \neq 0}}%
  ^{m=+\infty} \frac{h_m}{\addedall{|}m\addedall{|}^{1/2}}
  \exp\left[\icomp\left( 2\pi m \frac{\taunl}{\pernl} -\addedall{\sgnm}\frac{\pi}{4}\right)
    -(\addedall{\sgnm}\icomp+1)\left(\frac{\addedall{|}m\addedall{|}\pi}{\pernl}\right)^\shalf s \right],
\end{multline}
\addedall{where $\sgnm=\sgn{(m)}$}. It is interesting to
  note that the first order oscillatory solution $\coscinf$ depends on
  the boundary value of the zeroth-order smooth solution $\csm_0(0)$,
  which is to be determined in the next section.
% Equation
% (\ref{coscresult}) will be used to put the present model to the test
% by comparing it to a full numerical solution.
 
%=========================================================
\subsection{Second-order}
%=========================================================
The second-order oscillatory problem allows the determination of the
separation constant $\gsep_1$, which, as will be seen below, is enough
to solve the smooth problem up to order $\itpemun$. The calculation is
detailed in appendix \ref{annoscordre2} and yields
\begin{equation}
  \label{defgsep1}
  \gsep_1 = - \beta
    \left< 
      H(\taunl)\left[
        \csm_1(0,\tau)
      \right] 
    \right>_{\taunl}.
\end{equation}
The second-order oscillatory field $\cosc_2$ could also be obtained
analytically by using appendix \ref{annosc}, but is not required in
the present analysis.
%%%%%%%%%%%%%%%%%%%%%%%%%%%%%%%%%%%%%%%%%%%%%%%%%%%%%%%%%%
\section{The smooth problem}
\label{secsmooth}
%%%%%%%%%%%%%%%%%%%%%%%%%%%%%%%%%%%%%%%%%%%%%%%%%%%%%%%%%%
To treat the smooth problem, time is first rescaled by defining the
slow time-variable $\lambda = \tau / \pe$, and the smooth field $\csm$
is considered as a function of both fast and slow time-variables,
respectively $\tau$ and $\lambda$. The smooth equation
(\ref{eqsmooth}) reads, in the new variables
\begin{equation}
  \label{eqsmoothlambda}
  \dsurd{\csm}{\tau}+\ipemun\dsurd{\csm}{\lambda} 
  = \ipemun \dsurd{}{\sigma}
  \left[
    A(\sigma,\tau) \dsurd{\csm}{\sigma} + 
    \beta B(\sigma,\tau) \left(\csm + 1 \right)
  \right].
\end{equation}
The smooth field $\csm(\sigma,\tau,\lambda)$ is next expanded in the
small parameter $\itpemdemi$:
\begin{equation}
  \label{devcsm}
  \csm (\sigma, \tau, \lambda) = 
  \csm_0 (\sigma, \tau, \lambda) 
  + \ipemdemi \csm_1 (\sigma, \tau, \lambda)
  + \frac{1}{\pe} \csm_2 (\sigma, \tau, \lambda) + \dots,
\end{equation}
which, once introduced in equation (\ref{eqsmoothlambda}), yields a
hierarchy of equations in the small parameter $\itpemdemi$. As in
\cite{fyrillasszeri2}, the zeroth- and first-order smooth equations
read simply:
%
%\begin{equation}
%  \label{csm0tauindep}
%  \dsurd{\csm_0}{\tau} = 0 \Rightarrow \csm_0 (\sigma,\lambda) 
%\end{equation}
%                                
%and the zeroth and first order smooth equations
%are
%%
\begin{subeqnarray}
  \label{csm01tauindep}
  \dsurd{\csm_0}{\ds\tau} = 0 &\Rightarrow& \csm_0 (\sigma,\lambda) \\
  \dsurd{\csm_1}{\ds\tau} = 0 &\Rightarrow& \csm_1 (\sigma,\lambda)
\end{subeqnarray}
% \begin{equation}
%   \label{csm01tauindep}
%   \renewcommand{\arraystretch}{2}
% \left.
%   \begin{array}{l}
%   \dsurd{\csm_0}{\ds\tau} = 0 \Rightarrow \csm_0 (\sigma,\lambda) \\
%   \dsurd{\csm_1}{\ds\tau} = 0 \Rightarrow \csm_1 (\sigma,\lambda)
%   \end{array}
%  \right\},
% \end{equation}
%
which indicates that $\csm_0$ and $\csm_1$ vary with time only through
the slow time-scale $\lambda$.  The dependance of these two fields on
$\sigma$ and $\lambda$ can be obtained by writing the problems at
orders 2 and 3, and using a non-secularity condition. The smooth
boundary condition (\ref{clsmooth}) is written at each order by using
the expressions (\ref{defgsep0}) and (\ref{defgsep1}) of the
separation constants $G_0$ and $G_1$, and asymptotic solutions for
$\lambda\rightarrow\infty$ are sought. The technical details of the
calculation can be found in appendix \ref{annsmooth} \cite[see
also][]{fyrillasszeri1,fyrillasszeri2}. The resulting asymptotic
zeroth-order smooth field reads
\begin{equation}
  \label{solsmoothfinal}
  \czi(\sigma) = \exp\left[\beta\int_\sigma^\infty
%    \left<\ds\frac{4}{9}\ds\frac{\vp^2}{(3\sigma'+V)} \right>_\tau
%    \frac{\ddint\sigma'}{\left<(3\sigma'+V)^{4/3}\right>_\tau}
    \frac{\left<B(\sigma',\tau) \right>_{\tau}}
    {\left<A(\sigma',\tau) \right>_{\tau}}\:\ddint\sigma'
    \right] - 1,
\end{equation}
while the asymptotic first-order smooth field $\cui$ is zero, so that
equation (\ref{solsmoothfinal}) represents in fact the asymptotic
smooth solution up to order $1/\pe$.  

% The physical meaning of the asymptotic smooth solution may be
% understood by looking at equations (\ref{eqsminteg01}\,\textit{a,b}).
% \addedthree{The} average pressure diffusion flux (the $B$ term) is
% exactly balanced by the average Fick diffusion flux (the $A$ term),
% and therefore the smooth concentration field stays constant. The
% unsteady term in the smooth equations (\ref{eqsm01}\,\textit{a,b})
% represent the transitory non-equilibrium between the two average
% diffusion processes.
% Incidentally, equation
% (\ref{solsmoothfinal}) is the solution that one would have obtained by
% seeking a time-independant averaging naively the right-hand-side of equation (\ref{conservfin}).

%%%%%%%%%%%%%%%%%%%%%%%%%%%%%%%%%%%%%%%%%%%%%%%%%%%%%%%%%%
\section{Numerical results}
\label{secresults}
%%%%%%%%%%%%%%%%%%%%%%%%%%%%%%%%%%%%%%%%%%%%%%%%%%%%%%%%%%
%=========================================================
\subsection{Bubble dynamics}
% =========================================================
We will consider hereafter the case of an argon bubble in a mixture of
water and some other species at ambient temperature $T=298$ K. The
temporal evolution of the bubble radius is calculated by solving the
Keller-Miksis equation \cite[][]{kellermiksis, hilgenlohsebrenner96}:
\begin{multline}
  \label{kellerdim}
  \rr\frac{d^2 \rr}{dt^2} \left(1-\frac{1}{c}\frac{d\rr}{dt} \right) 
  + \troisdemi\left(\frac{d\rr}{dt} \right)^2 
  \left(1-\frac{1}{3c}\frac{d\rr}{dt} \right) = \\
  \frac{1}{\rho}\left\{
    \left(1+\frac{1}{c}\frac{d\rr}{dt}\right)
    \left[\ppg  - p_0(1-P\cos \omega t)  \right]
    + \frac{\rr}{c} \frac{d \ppg}{dt} - \frac{2\sigma}{\rr}
    -\frac{4\mu}{\rr}\frac{d\rr}{dt}
  \right\},
\end{multline}
where $\rr$ is the bubble radius, $\ppg$ the gas pressure in the
bubble, assumed homogeneous, $c=1500$\mpersec, $\rho=998$\kgpermcube,
$\mu=10^{-3}$\viscunit are respectively the sound velocity, density
and dynamic viscosity of water, $p_0 = 101325$\presunitbis the
pressure in the liquid at rest, $\sigma = 0.072$\tsurfunit the
water-gas surface tension, $P$ the dimensionless driving pressure
amplitude and $\omega$ the angular driving frequency.

Two different models can be used for the bubble interior. The
\addedthree{first} assumes an isothermal behaviour and a van der Waals
equation of state, so that the bubble internal pressure is
\begin{equation}
  \label{vanderwaals}
  \ppg =\left(p_0 + \frac{2\sigma}{\rz} \right)
  \left(\frac{\rz^3-h^3}{{\rr^3-h^3}} \right),
\end{equation}
where $\rz$ is the ambient radius of the bubble and $h$ the van der
Waals hard-core radius. A refined model accounting for water
evaporation at the bubble wall and temperature gradients in the bubble
was also used. The details of the model can be found elsewhere
\cite[][]{toegel2000, storeyszeri2001}.  It is known that accounting
for such effects reduces the violence of the collapse and may
therefore influence the segregation process investigated in this
paper, as will be seen below.

In the following sections, equation (\ref{kellerdim}) will be solved
for various sets of parameters $\omega$, $P$ and $\rz$,
over a number of periods sufficiently large to get steady-state
oscillations. The corresponding bubble volume and its
time-derivatives on the last period are stored in tables, from which
the time and space dependent coefficients $A(\sigma,\tau)$ and
$B(\sigma,\tau)$ can be calculated by equations
(\ref{defAB}\,\textit{a,b}) when needed.
%=========================================================
\subsection{Comparison with full simulation}
\label{seccomparnumeric}
%=========================================================
In order to check the validity of the approximation obtained from the
splitting method, numerical simulations of the full
convection-diffusion problem (\ref{conservfin})-(\ref{boundinlag})
have been performed, with the help of the FEMLAB software. The present
set of equations is recast without further difficulty in the canonical
coefficient form of partial differential equations allowed in FEMLAB.
The interval $[0, \infty]$ was mapped to $[0, 1]$ by using
the variable change $x = 1/(\sigma+1)$, the interval $[0,1]$ was
non-uniformly meshed to trap the boundary layer near the bubble wall,
and mesh convergence studies were performed to ensure good accuracy of
the result.

In order to test the analytical approximation obtained in the
preceding section, we first recall that the analytical method yields
the concentration field as
\[
\cad(\sigma,\tau) = \csm_0(\sigma,\lambda) 
+ \ipemdemi \csm_1(\sigma,\lambda) 
+ \ipemdemi \cosc_1(\sigma,\tau) + O\left(\ipemun \right),
\]
since $\cosc_0 = 0$. For large times ($\lambda\rightarrow\infty$),
$\csm_0$ reaches its asymptotic limit $\czi$ and as shown above,
$\csm_1$ vanishes. The oscillatory field $\cosc_1$ should reach its
asymptotic value (\ref{coscresult}) in a few periods, and therefore
one should have
\begin{equation}
  \label{comparsimanaltot}
  \cad(\sigma,\tau) \underset{\lambda\rightarrow\infty}{\sim}
  \czi(\sigma) 
  + \ipemdemi \coscinf(\sigma,\tau) + O\left(\ipemun \right)  .
\end{equation}
Further averaging on time $\taunl$ over one period, we get
\begin{equation}
  \label{comparsimanalsm}
\left<\cad(\sigma,\tau) \right>_{\taunl} 
 \underset{\lambda\rightarrow\infty}{\sim}
  \czi(\sigma) + O\left(\ipemun\right),
\end{equation}
since from equation (\ref{coscresult}), $\coscinf$ has a null
$\taunl$-average.

Both equations (\ref{comparsimanaltot}) and (\ref{comparsimanalsm})
were checked against direct numerical simulation for an argon bubble
of ambient radius $R_0=4$\microns{} driven by pressure fields of
\addedtwo{dimensionless} amplitudes $P=$ 0.3, 0.6 and 0.8 and
frequency 26.5 kHz.  Since our aim is to check the analytical model
against a numerical result, we take an arbitrary value
$\beta=-10^{-5}$ rather than specifying the species A mixed with
water. In order to reach the limit $\lambda\rightarrow\infty$
numerically, the final time of the simulation was chosen sufficiently
large so that the system nearly reaches its steady state. The analysis
of the smooth problem shows that its steady state should be obtained
within a number of periods of the order of $\pe$. We therefore chose
arbitrarily \addedtwo{$\pe=100$ and $\pe=500$} in order to get reasonable
simulation times. We found in our examples that no noticeable change
from one period to the following one could be observed after about
$2\pe$ periods.  The last oscillation period of the concentration
field $\cnum(\sigma,\tau)$ obtained numerically was stored, the
nonlinear time $\taunl$ was calculated, and the nonlinear-average
$\left<\cnum(\sigma,\tau)\right>_\taunl$ was calculated over one
period at each spatial point $\sigma$. The smooth concentration field
$\czi(\sigma)$ was evaluated by calculating the integral in equation
(\ref{solsmoothfinal}) with a Gauss-Jacobi method \cite[see][appendix
B for details]{louisnardgomez2003}. The asymptotic oscillatory field
was calculated from equation (\ref{coscresult}), after evaluating the
Fourier coefficients $h_m$ of $H(\taunl)$ by a fast Fourier-transform.

\begin{figure}
  \centerline{\includegraphics[width=0.9\linewidth]{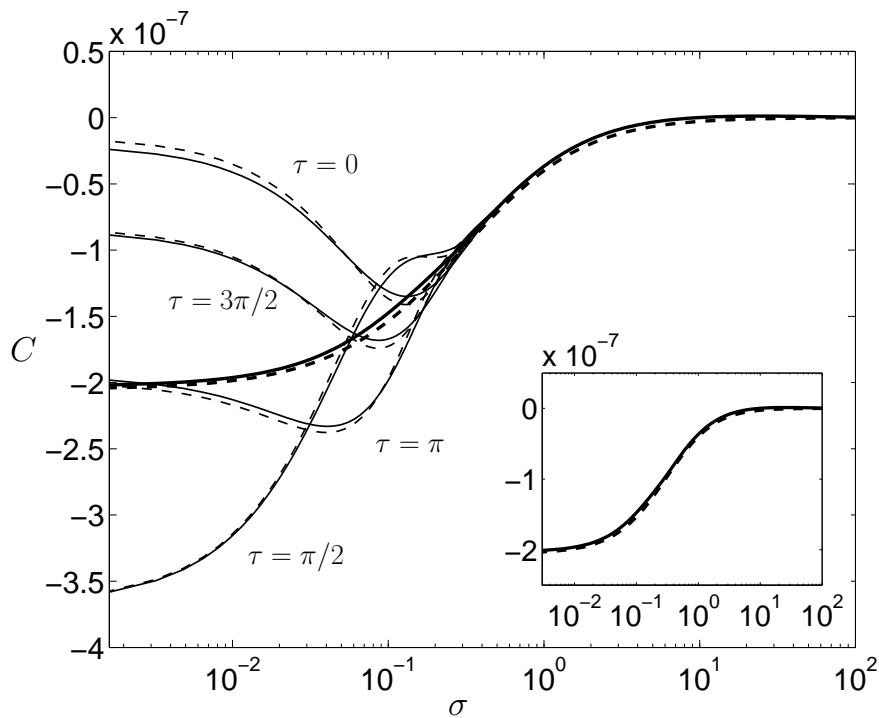}}
  \caption{Comparison between the full numerical solution and the
    analytical approximation for a 4\microns{} argon bubble in water in
    a 26.5 kHz acoustic field of amplitude $P=0.6$. The P\'eclet number
    \addedtwo{is 500} and the parameter $\beta$ is $-10^{-5}$.  Thin solid lines:
    concentration profiles $\cnum(\sigma,\tau)$ obtained by numerical
    simulation at different phases of the bubble oscillation. Thin
    dashed line: analytical predictions $\czi(\sigma) +
    \pe^{-1/2}\coscinf(\sigma,\tau)$.  Thick solid line: nonlinear
    numerical average of the numerical profile over one period.  Thick
    dashed line: asymptotic zeroth-order smooth concentration profile
    $\czi(\sigma)$. The inset shows a more detailed comparison between
    the numerical average and the smooth solution.}
  \label{verylow}
\end{figure}

Figure~\ref{verylow} shows typical concentration profiles results for
a 4 $\mu$m argon bubble in water, driven by an oscillatory pressure of
0.6 bar amplitude and 26.5 kHz frequency, \addedtwo{and $\pe=500$}.
The dashed lines represent the analytical predictions and the solid
ones are the numerical \addedall{results}.  The total concentration
profile (thin lines) is drawn at four distinct phases of the acoustic
period in order to check equation
(\ref{comparsimanaltot})\addedthree{. It is seen} that the analytical
predictions are in excellent agreement with the numerical result.  We
also display in figure~\ref{verylow} the average
$\left<\cnum(\sigma,\tau)\right>_\taunl$ (thick solid line) along with
the analytical prediction $\czi(\sigma)$ (thick dashed
line)\addedthree{. It can} be seen that both quantities are in
excellent agreement (see the magnification in the inset) and we
conclude that equation (\ref{comparsimanalsm}) is fulfilled.
\addedtwo{Besides, it is expected that the analytical approximation
  would progressively break as the asymptotic parameter $\itpemdemi$
  increases.  Calculations with a smaller P\'eclet number ($\pe=100$,
  not presented here), show that this is indeed the case, and yielded
  a maximum relative error on the oscillatory field amounting to
  11\%.}

\begin{figure}
  \centerline{\includegraphics[width=0.9\linewidth]{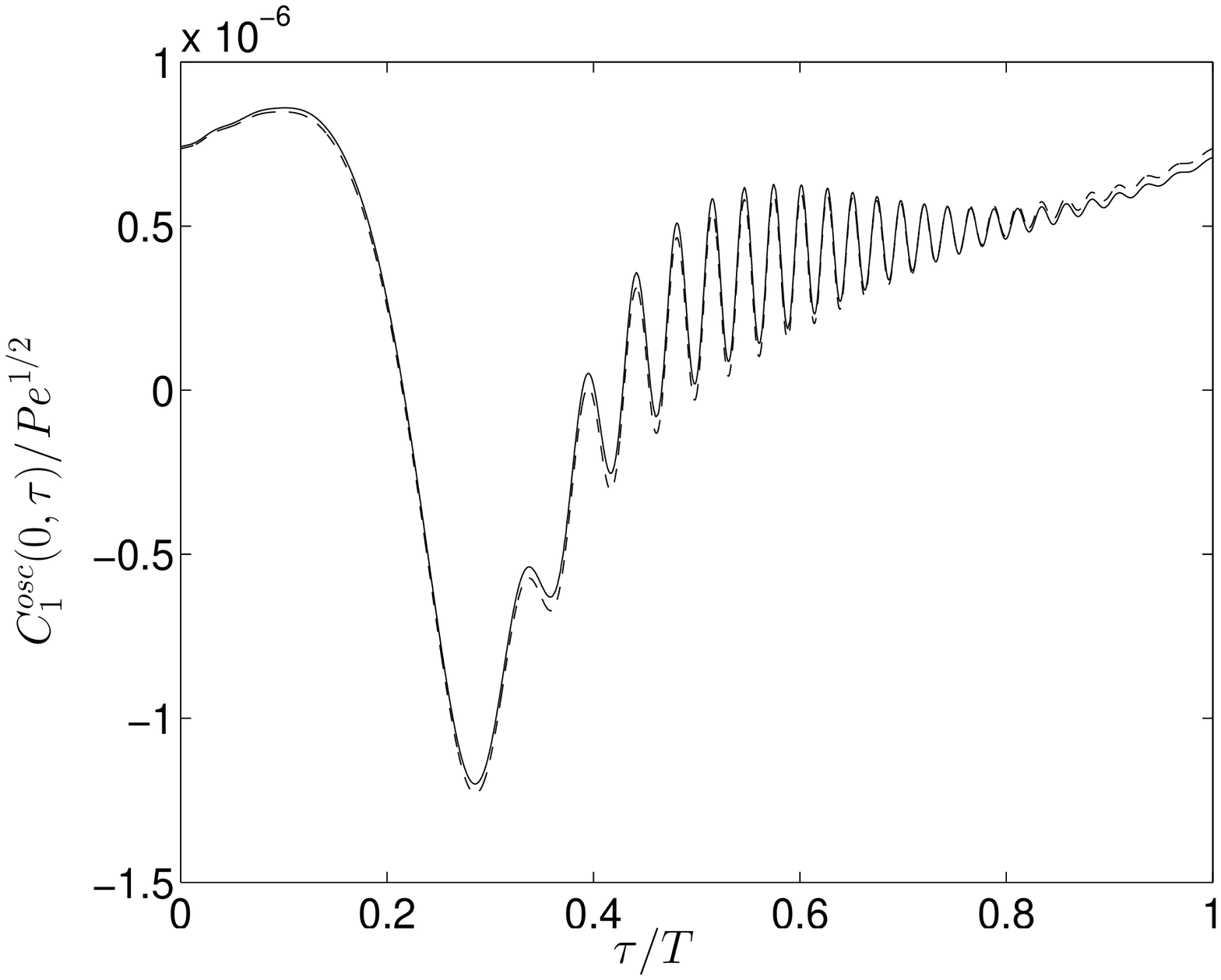}}
  \caption{Comparison between the analytical oscillatory concentration
    (dashed line) $\pe^{-1/2}\coscinf(0,\tau)$ at the bubble wall and
    the numerical solution $\cnum(0,\tau) - \czi(0)$ (solid line) for
    a 4 $\mu$m argon bubble driven at $P = 0.8$ and 26.5 kHz. The
    P\'eclet number is 100 and the parameter $\beta$ is $-10^{-5}$.}
  \label{comparosc}
\end{figure}

Another validation of the model can be seen in figure \ref{comparosc},
which compares the oscillatory part of the numerical solution at the
bubble wall $\cnum(0,\tau)-\czi(0)$ to the analytical solution
$\pe^{-1/2}\coscinf(0,\tau)$ over one period of oscillation, for a
driving pressure of amplitude $P=0.8$\addedtwo{, and $\pe=100$}: here
again, the two results are in excellent agreement.

%=========================================================
\subsection{Parameter-space exploration}
%=========================================================
The validation of the analytical model being achieved, we now turn to
investigate how the smooth and oscillatory parts vary with the bubble
parameters $(R_0,P,\omega)$. In order to get an immediate view of the
magnitude of the segregation process, we will focus on the values of
the two fields at the bubble wall.

%---------------------------------------------------------
\subsubsection{Smooth part}
%---------------------------------------------------------
The smooth concentration at the bubble wall $\czi(0)$ is obtained by
setting $\sigma=0$ in equation (\ref{solsmoothfinal}):
\begin{equation}
  \label{csmoothwall}
\czi(0) = \exp(\beta I) - 1,
\end{equation}
where 
\begin{equation}
  \label{defIsmooth}
  I = \int_0^\infty
  \frac{\left<B(\sigma',\tau) \right>_{\tau}}
  {\left<A(\sigma',\tau) \right>_{\tau}}\:\ddint\sigma'.
\end{equation}
The value of integral $I$ depends only on the bubble dynamics, and in
order to get a picture independent of the choice of a specific
mixture, but containing all the bubble data, we  use the
definition (\ref{defbetadbis}) of the parameter $\beta$ to obtain
\begin{equation*}
  \czi(0) = \exp(\beta_m \rz^2\omega^2 I) - 1,
\end{equation*}
where $\beta_m$, defined by equation (\ref{defbetam}) depends only on
the mixture considered. Thus, the value of $\rz^2\omega^2 I$ will be
calculated for various bubble parameters and $\czi(0)$ can then be
easily deduced for a specific mixture.

\begin{figure}
  \centerline{\includegraphics[width=0.9\linewidth]{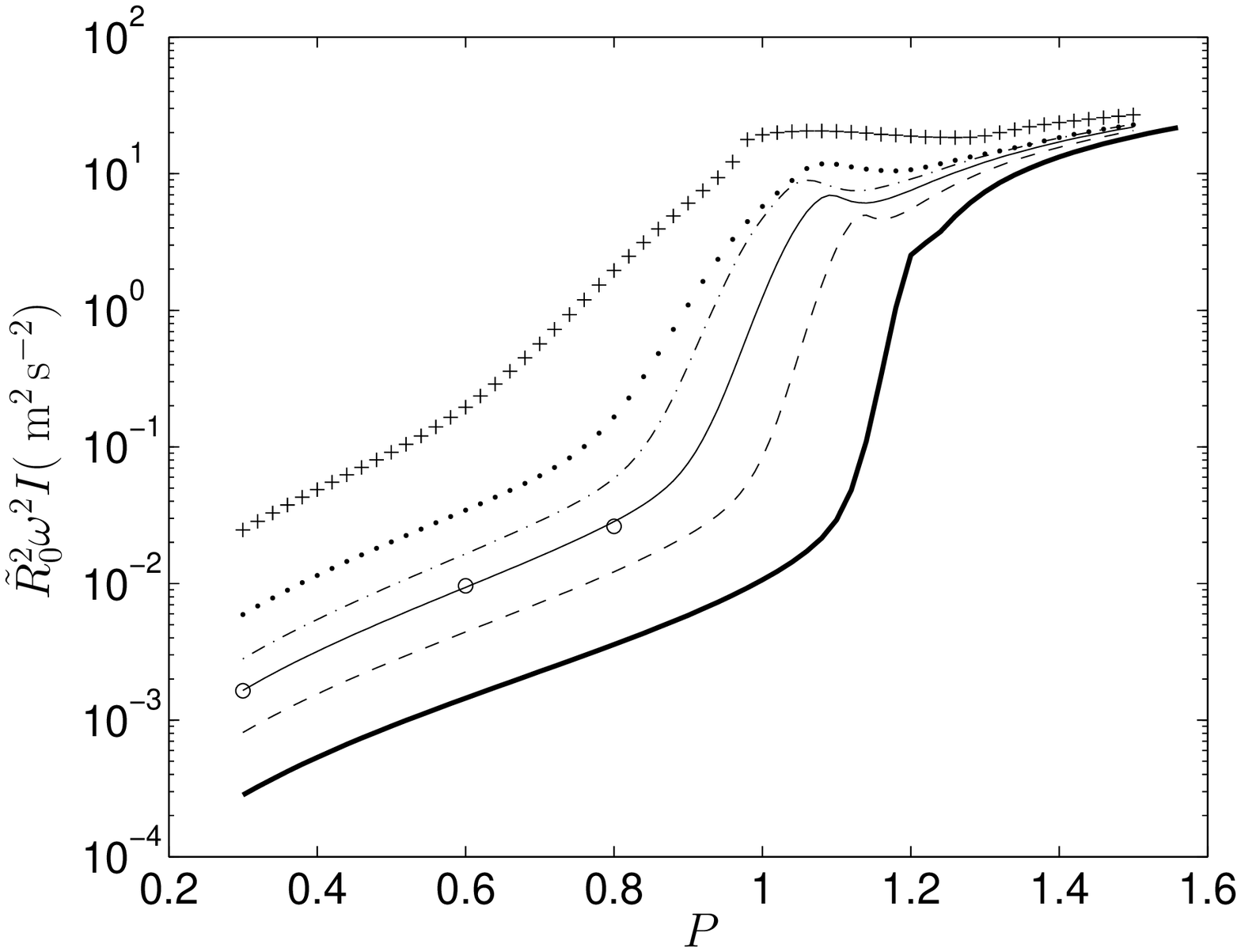}}
  \caption{Evolution of $\rz^2\omega^2 I$ with driving pressure, from
    equation (\ref{defIsmooth}). The four bottom curves are calculated
    with $f =$26.5 kHz, for ambient radii $\rz=$2\microns{} (thick solid
    line), 3\microns{} (dashed line), 4\microns{} (thin solid line),
    5\microns{} (dash-dotted line). The two top curves are calculated
    for $\rz=$4\microns{} and respectively with $f =$50 kHz (dotted
    line) and $f=$100 kHz (+ signs).  The three circles represents the
    results obtained by FEMLAB full simulations for 4 $\mu$m bubbles
    driven respectively by pressure fields of 0.3, 0.6 and 0.8 driving
    pressure.}
  \label{figsmoothpar}
\end{figure}

Figure \ref{figsmoothpar} represents $\rz^2\omega^2 I$ as a function
of the driving pressure, for different ambient radii and different
frequencies.  The four bottom
curves are calculated for a frequency of 26.5 kHz, for bubble ambient
radii ranging from 2\microns{} to 5\microns{}. It can be seen that
$\rz^2\omega^2 I$ increases with $\rz$ in the range considered. The
three circles represent the value obtained from FEMLAB direct
simulations, showing again the good agreement with analytical results.
The two top curves are calculated for a 4\microns{} bubble excited
respectively at 50 kHz (dotted line) and 100 kHz (+ signs): it can be
seen that the mean segregation process increases markedly with
frequency for small driving pressures, but that all curves merge for
high driving pressures.

It can be noted that in all cases, a marked increase of $\rz^2\omega^2
I$ occurs near $P=1$ which is approximately the Blake threshold
\cite[][]{akhatovgumerov97, hilgenbrennergrosslohse98,
  louisnardgomez2003}. Above this driving pressure value, the bubble
dynamics becomes inertially driven, yielding large time-variations of
$V(t)$ and its time-derivatives, and therefore large values of the
integrand in equation (\ref{defIsmooth}).

%=========================================================
\subsection{Parameter-space exploration : oscillatory part}
\begin{figure}
  \centerline{\includegraphics[width=0.9\linewidth]{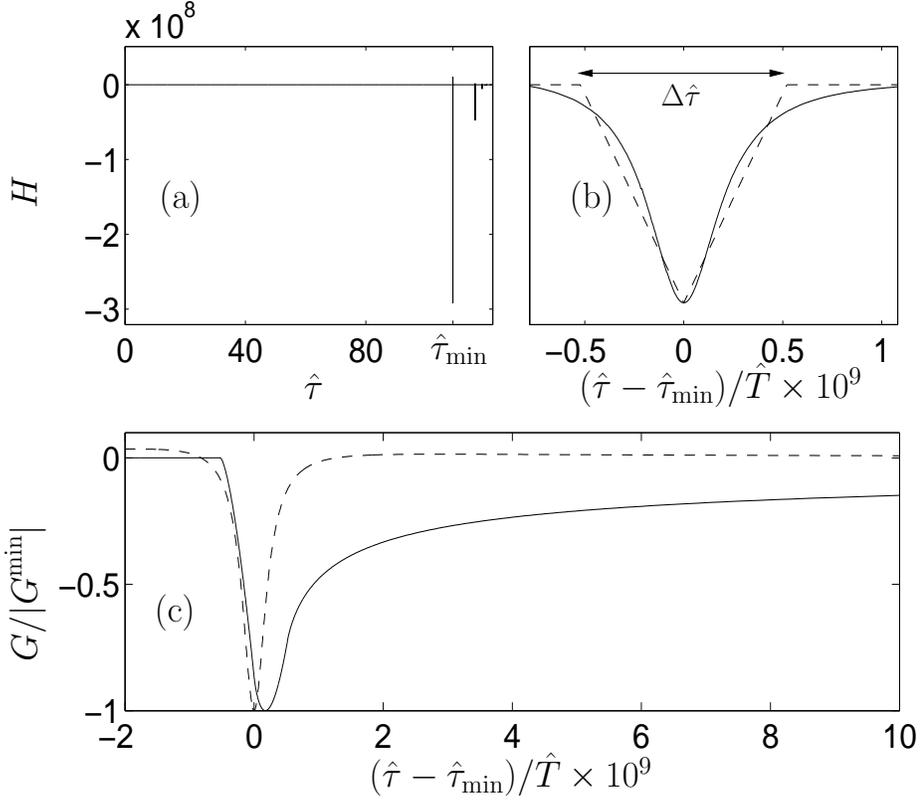}}
  \caption{(\textit{a}) Time evolution of function $H(\taunl)$ for a 4
    $\mu$m bubble driven at $P=1.1$ and 26.5 kHz. The nonlinear period
    is 122 in this case. The negative peaks corresponds to the huge
    positive values taken by $\vpp$ at the main collapse and
    subsequent afterbounces.  (\textit{b}) Zoom on the most negative
    peak of $H(\taunl)$ (solid line) translating the origin of
    abscissas to the location of this peak.  It is seen that the width
    of the peak is 9 orders of magnitude smaller than the nonlinear
    period; the dashed line is the approximation of $H(\taunl)$
    defined by equation (\ref{hfit}) (\textit{c}) Solid line: Shape of
    function $\gser(\taunl)$ given by equation (\ref{defgser}).
    Dashed line: shape of function $H(\taunl)$. }
  \label{figshapeGH}
\end{figure}

%=========================================================
Neglecting terms of order $O(\pe^{-1})$, the oscillatory concentration
at the bubble wall $\cosc(0,\taunl)$ reduces to
$\coscinf(0,\taunl)/\pemdemi$. Evaluating equation (\ref{coscresult})
at $s=0$, we get
\begin{equation}
  \label{coscwall}
  \cosc(0,\taunl) =
  \frac{\beta}{\pemdemi} \left[ \csm_0(0) + 1 \right] \gser(\taunl),
\end{equation}
where
\begin{equation}
  \label{defgser}
  \gser(\taunl) = \left(\frac{\pernl}{2\pi} \right)^{\shalf} 
  \sum_{%
    \substack{
      m = -\infty \\
      m \neq 0}}^{m=+\infty} \frac{h_m}{\addedall{|}m\addedall{|}^{1/2}}
  \exp\left[\icomp\left( 2\pi m \frac{\taunl}{\pernl}
      -\addedall{\sgnm}\frac{\pi}{4}\right)
  \right].
\end{equation}
Using equation (\ref{defbetadbis}) to express the factor
$\beta/\pemdemi$ in terms of the dimensional parameters, equation
(\ref{coscwall}) becomes
\begin{equation}
  \label{betpemdemi}
  \coscinf(0,\taunl) =
  \beta_m D^{1/2}\rz\omega^{3/2} \left[ \csm_0(0) + 1 \right] \gser(\taunl).
\end{equation}
In order to identify the contribution of the bubble oscillations
independently from the choice of a specific mixture, the quantity
$\rz\omega^{3/2} \gser(\taunl)$ must be calculated for various bubble
parameters. For small driving pressure, $\gser(\taunl)$ can be
evaluated by summing the series (\ref{defgser}) without any specific
problem, as was done in section \S\ref{seccomparnumeric}.  However, for
driving pressures high enough to yield inertial cavitation, evaluation
of $\gser(\taunl)$ is subject to a technical difficulty linked to the
shape of function $H(\taunl) = -2{\rpp}/{R^2}$, as shown in figure
\ref{figshapeGH}(\textit{a})\addedthree{. Owing to} the huge outward acceleration
of the liquid at the end of the bubble collapse, $H(\taunl)$ looks
like a series of negative Dirac distributions, the most important
being located at the main collapse, and the other ones at each
secondary collapse between the bubble afterbounces.  From the singular
shape of function $H(\taunl)$, it is expected that its Fourier
spectrum (the coefficients $h_m$ in the series (\ref{devfourierH}))
spans over a wide frequency range.  Therefore series (\ref{defgser})
converges very slowly, thus forbidding any numerical estimation. This
is illustrated in figure \ref{figshapeGH}(\textit{b}), which shows a
magnification of the most negative peak of $H(\taunl)$\addedthree{. It is seen}
that the width of the peak is less than 9 orders of magnitude the
nonlinear period of oscillation, so that one should sum more than
$10^9$ terms in the series to obtain an acceptable result~!

We therefore used the following trick: let's denote $\taunlmin$ the time
at which $H(\taunl)$ reaches its highest negative peak amplitude $\hmin$.
We fit $H$ by a negative tooth function of amplitude $\hmin$ and width
$\Delta \taunl$: 
\begin{equation}
  \label{hfit}
  H(\taunl) \simeq \left\{
    \begin{array}{ll}
      \hmin
      \left(1-\left|\displaystyle\frac{\taunl-\taunlmin}{\Delta\taunl} \right|
      \right), 
      & \taunl\in [\taunlmin- \Delta\taunl, \taunlmin+ \Delta\taunl ]
      \\
      0, & \text{elsewhere}
    \end{array}
    \right..
\end{equation}
where $\Delta\taunl$ is determined in such a way that the real and
fitted peaks have the same integral over the interval $[\taunl_1,
\taunl_2]$, where $\taunl_1$ and $\taunl_2$ are the locations of the
zeros of $H(\taunl)$ at each side of $\taunlmin$:
\begin{equation}
  \label{defdeltatau}
  \Delta \taunl = \frac{1}{\hmin}\int_{\taunl_1}^{\taunl_2} H(\taunl)
  \:\ddint \taunl.
\end{equation}
$\Delta \taunl$ represents physically the characteristic time (in
nonlinear form) of the bubble rebound at the end of the collapse.
Figure \ref{figshapeGH}(\textit{b}) shows the original function
$H(\taunl)$ (solid line) compared to the approximation obtained by
equation (\ref{hfit}) (dotted line).

The Fourier coefficients of the tooth function can be easily calculated and
introduced in equation (\ref{defgser}) to calculate $G(\taunl)$. This is
done in appendix \ref{annpeak} and the following approximation of $G(\taunl)$
is obtained:
\begin{equation}
  \label{defgappser}
  \gapp(\taunl) =   \left(\frac{\pernl}{2\pi}
  \right)^{1/2}\hmin\frac{\pernl}{\pi^2\Delta \taunl}
  \sum_{m=1}^{m=+\infty}\frac{1}{m^{5/2}} 
  \sin^2 \left(m\pi\frac{\Delta \taunl}{\pernl} \right)
  \cos\left[2m\pi\frac{\taunl-\taunlmin}{\pernl}-\frac{\pi}{4} \right].
\end{equation}
For very small $\Delta \taunl$, which is the case for inertial cavitation,
this series is as difficult to calculate as the original one in
equation (\ref{defgser}). However, a good approximation of
$\gapp(\taunl)$ can be found and is detailed in appendix \ref{annpeak}
(see equations (\ref{gappann}) and (\ref{fapprox})). Figure
\ref{figshapeGH}(\textit{c}) shows the typical shape of
$\gapp(\taunl)$: it decreases rapidly down to a minimum located slightly
after the minimum of $H$ and then slowly relaxes to 0. We first
restrain our primary interest to the extremal value attained by
$\coscinf(0,\taunl)$ over one period, so that only the minimum value
of $\gapp(\taunl)$ is needed. It is shown in appendix \ref{annpeak}
that an excellent estimate of this minimum is
\begin{equation}
  \label{defgappmin}
  \gappmin = \frac{8\gamundemi}{3\sqrt{3}\pi}\hmin\Delta\taunl^{1/2}.
\end{equation}

\begin{figure}
  \centerline{\includegraphics[width=0.9\linewidth]{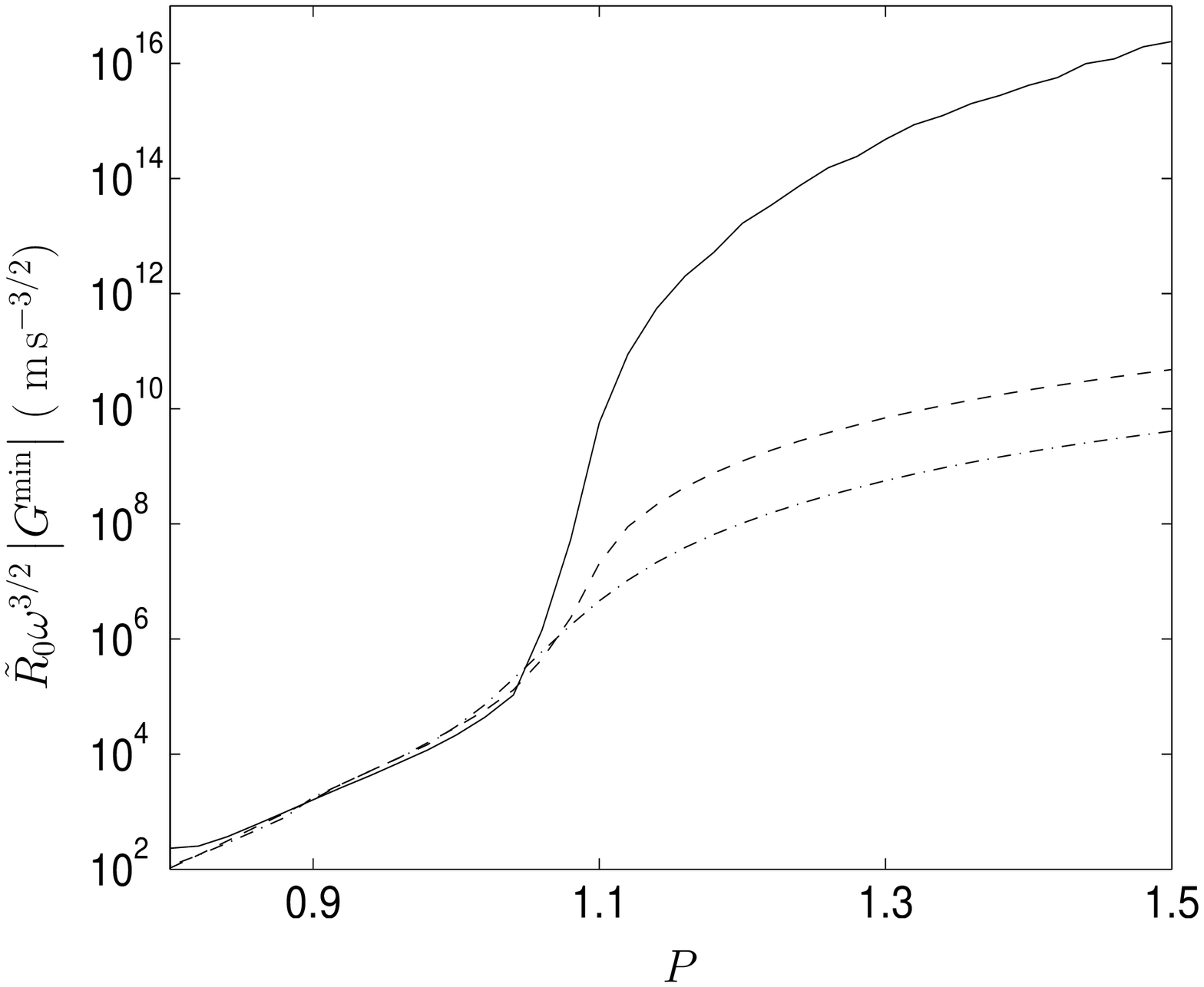}}
  \caption{Solid line: evolution of $\rz\omega^{3/2}\left|\gmin
    \right|$ calculated by summing the series in equation
    (\ref{defgser}) from a $2^{15}$ points FFT of $H(\taunl)$. Dashed
    line: $\rz\omega^{3/2}\left|\gappmin \right|$ calculated from
    (\ref{defgappmin}). Both curves are obtained for a 4\microns{}
    argon-bubble excited at 26.5 kHz assuming an isothermal gas
    behaviour.  The dash-dotted line also represents the evolution of
    $\rz\omega^{3/2}\left|\gappmin \right|$, but calculated with the
    refined model of the bubble interior. The values obtained are
    about one order of magnitude smaller than with the isothermal model.}
  \label{figallcosc}
\end{figure}

The solid line in figure \ref{figallcosc} displays the evolution of
$\rz\omega^{3/2}\left|\gmin \right|$ obtained by summing directly
series (\ref{defgser}) along with a 2$^{15}$ points FFT of
$H(\taunl)$, while the dashed line represents
$\rz\omega^{3/2}\left|\gappmin \right|$ calculated from the
approximate equation (\ref{defgappmin}) for a 4\microns{} argon-bubble
oscillating at 26.5 kHz in water.  It is seen that both results are in
agreement up to about $P=1.05$ ( which corresponds approximately to
the Blake threshold) and that they markedly diverge above the
threshold, which \addedthree{demonstrates} that for inertial motion of the bubble,
$H(\taunl)$ becomes too sharp to be correctly represented by a
reasonable Fourier expansion.  Therefore, in the inertial regime, the
approximate equation (\ref{defgappmin}) must be used to calculate
$\gmin$.

The dash-dotted line in figure \ref{figallcosc} also displays the
value of $\rz\omega^{3/2}\left|\gappmin \right|$ calculated from
equation (\ref{defgappmin}) but with a refined bubble interior model,
taking into account heat transport and water condensation/evaporation
at the bubble wall \cite[][]{toegel2000,storeyszeri2001}\addedthree{.
  At low} driving pressures, the results are comparable, but above the
Blake threshold, the refined model predicts values lower
\addedthree{by} one order of magnitude. Such a result could be
\addedthree{expected} since it is known that taking into account heat
transport in the bubble interior yields a less violent collapse than
with the isothermal model, and therefore decreases the amplitude of
function $H$. Similar conclusions have been drawn for other bubble
phenomena directly linked to the violence of the collapse, such as
Rayleigh--Taylor shape instabilities \cite[][]{linstoreyszeri2002}.
Since the refined model is believed to be more realistic than the
isothermal one, it will be used in every result presented hereafter.

\begin{figure}
  \centerline{\includegraphics[width=0.9\linewidth]{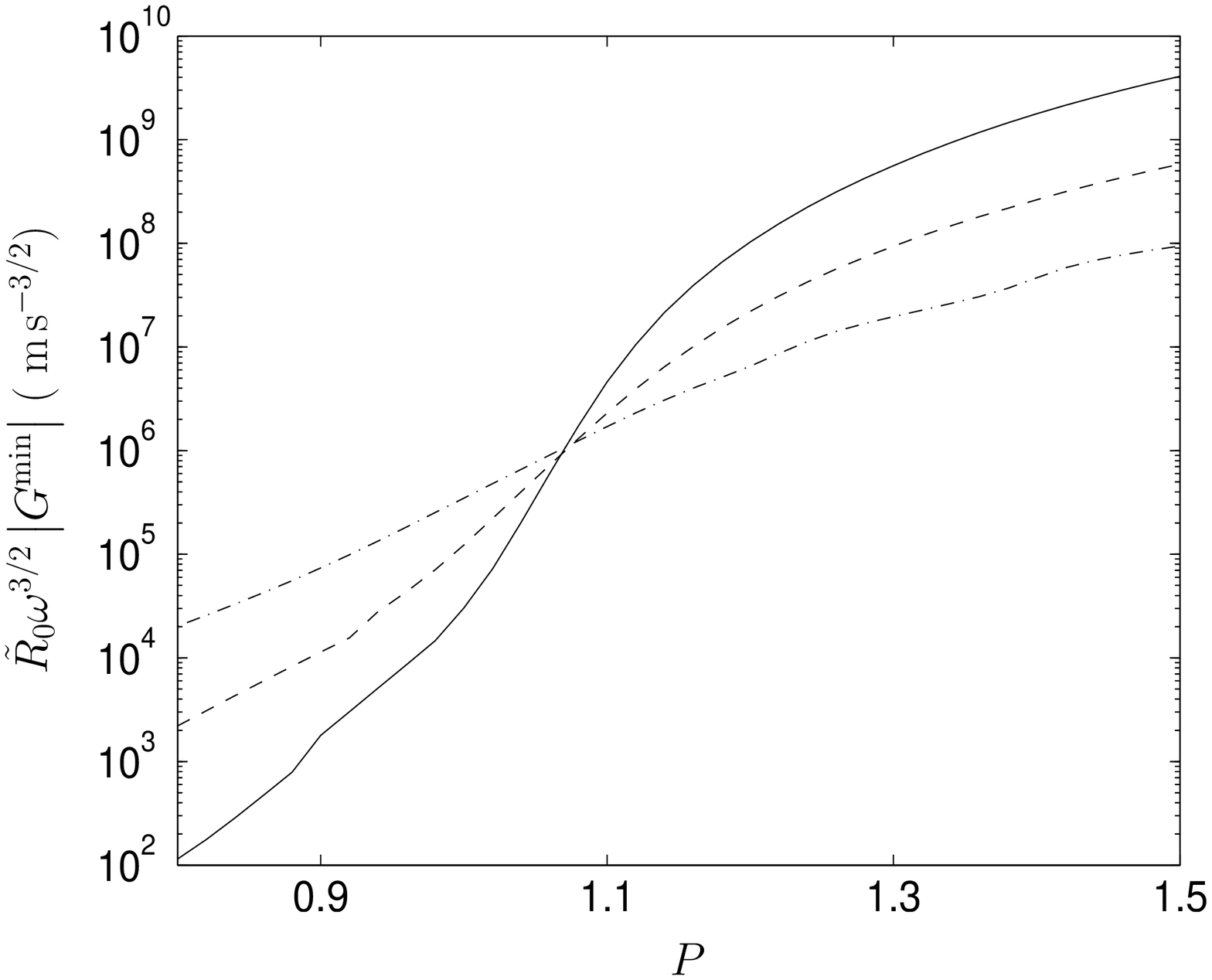}}
  \caption{Evolution of $\rz\omega^{3/2}\left|\gappmin \right|$
    calculated from (\ref{defgappmin}), for a 4\microns{} argon-bubble
    excited at 26.5 kHz (solid line), 50 kHz (dashed line) and 100 kHz
    (dash-dotted line). The bubble interior refined model was used in
    all cases.}
  \label{figcomparfreq}
\end{figure}

Figure \ref{figcomparfreq} displays the influence of frequency on the
oscillatory field. It is seen that as frequency increases, we get a
stronger effect at low amplitude but, for high amplitudes, increasing
the frequency reduces the oscillatory segregation effect, despite the
$\omega^{3/2}$ scaling law. This can be easily explained by the fact
that increasing the frequency limits the expansion phase of the bubble
in the inertial regime, which in turn reduces the violence of the
collapse, and therefore the peak value attained by the $H$ function.

\begin{figure}
  \centerline{\includegraphics[width=0.9\linewidth]{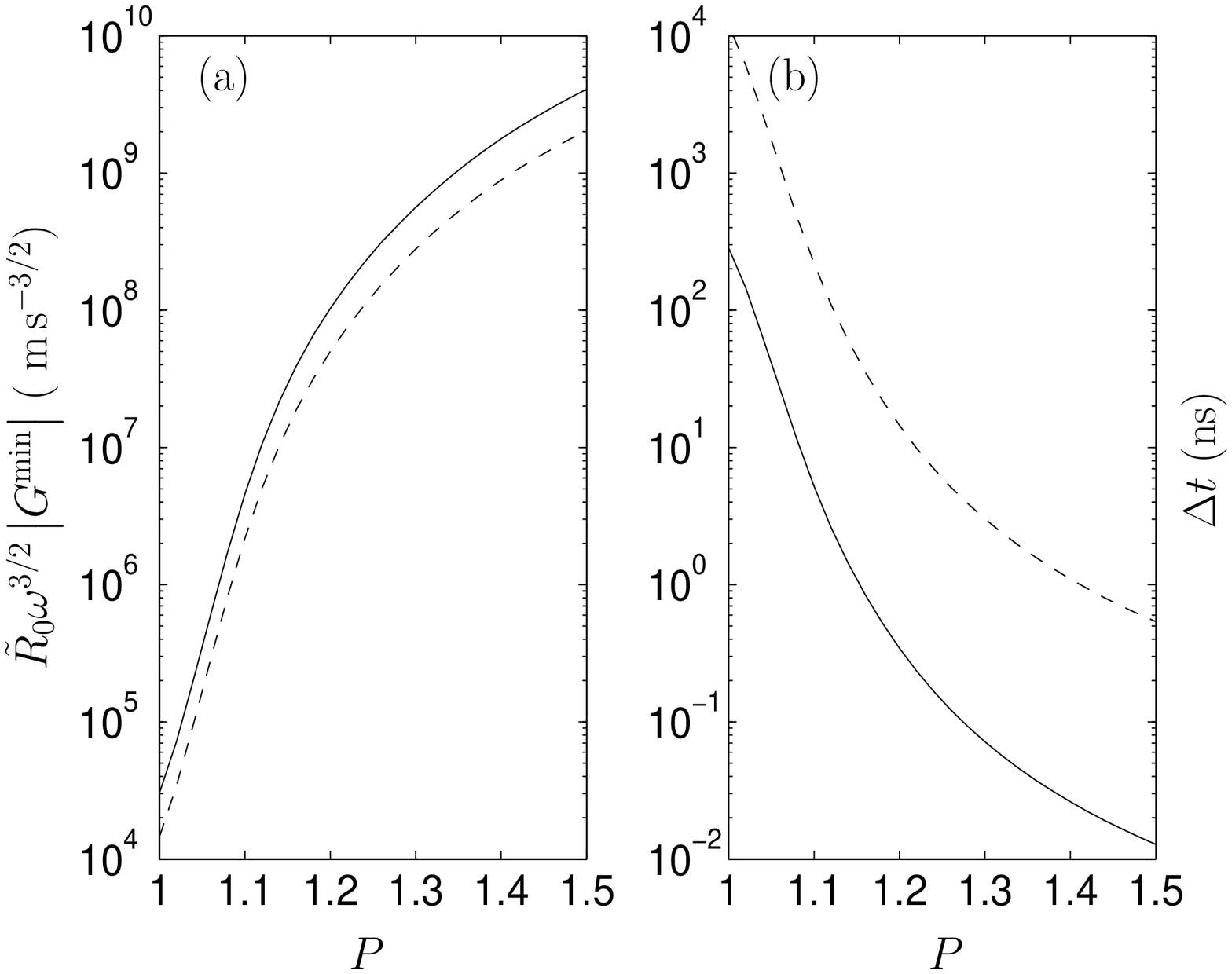}}
  \caption{(\textit{a}) Comparison of $\rz\omega^{3/2}\left|\gappmin
    \right|$ evaluated from equation (\ref{defgappmin}) (solid line)
    and from equation (\ref{gappminrapid}) (dashed line) for a
    4\microns{} argon-bubble excited at 26.5 kHz. (\textit{b})
    Characteristic time $\Delta t$ of the bubble rebound for the same
    bubble, calculated from equation (\ref{deltat}) (solid line).  The
    dashed line represents 42$\Delta t$ which is the time necessary
    for the oscillatory segregation to reach one tenth of its maximum
    value.}
  \label{figdeltatau}
\end{figure}

Finally, a more practical sense can be given to the time-interval
$\Delta\taunl$ appearing in equation (\ref{defgappmin}): since
$H(\taunl)=-2\rpp/R^2$, and using the definition (\ref{deftaunl}) of
the nonlinear time, equation (\ref{defdeltatau}) can also be expressed
as
\begin{equation*}
  \Delta \taunl = \frac{-2}{\hmin}\int_{\tau_1}^{\tau_2} R^2(\tau)\rpp(\tau)
  \:\ddint \tau.  
\end{equation*}
Times $\tau_1$ and $\tau_2$ are located respectively closely
before and closely after the time at which the bubble reaches its
minimum radius. Therefore $R$ stays close to $\rmin$ in the interval
$[\tau_1,\tau_2]$, so that $\hmin\simeq-2\rppmax/\rmin^2$. Therefore:
\begin{equation*}
  \Delta \taunl \simeq \frac{\rmin^4(\rp(\tau_2)-\rp(\tau_1))}{\rppmax},
\end{equation*}
and since by definition $\tau_1$ and $\tau_2$ are the zeros of $\rpp$,
$\rp(\tau_1)$ and $\rp(\tau_2)$ are the minimum and maximum 
bubble velocities attained before and after the rebound respectively,
which are in fact the minimum and maximum velocities of the bubble
over one acoustic period. Thus:
\begin{equation}
  \label{defdeltataubis}
  \Delta \taunl \simeq {\rmin^4}\frac{\rpmax-\rpmin}{\rppmax}.
\end{equation}
Injecting this value in equation (\ref{defgappmin}), and setting
$\hmin\simeq -2\rppmax/\rmin^2$, we get
\begin{equation}
  \label{gappminrapid}
    \gappmin \simeq -
    \frac{8\gamundemi}{3\sqrt{3}\pi}\left[\rppmax(\rpmax-\rpmin) \right]^{1/2},
\end{equation}
which can easily be evaluated once the bubble dynamics is
known. Figure \ref{figdeltatau}(\textit{a}) shows that equation
(\ref{gappminrapid}) gives a reasonable approximation of $\gappmin$. 

Apart from the minimum value reached by $G$, it is also of interest to
obtain an order of magnitude of the relaxation time of $G$ (see solid
line figure \ref{figshapeGH}\textit{c}). It is shown in appendix
\ref{annpeak} that $G$ reaches one tenth of its minimum value after a
relaxation time of $42\Delta\taunl$ past $\taunlmin$.  Evaluating
$\Delta\taunl$ from equation (\ref{defdeltataubis}), a ready-to-use
estimate of the oscillatory segregation duration can be obtained. The
dimensional rebound time $\Delta t$ corresponding to $\Delta\taunl$
can be obtained by first converting the latter in linear time by
$\Delta\tau\simeq\Delta\taunl/\rmin^4$ and setting $\Delta t = \Delta
\tau/\omega$. We obtain:
\begin{equation}
  \label{deltat}
  \Delta t \simeq \frac{1}{\omega}\frac{\rpmax-\rpmin}{\rppmax}.
\end{equation}
Figure \ref{figdeltatau}(\textit{b}) displays the dimensional rebound
characteristic time $\Delta t$ in ns (solid line) for a 4\microns{}
argon bubble at 26.5 kHz: it rapidly drops from about 300 ns for $P=1$
to 10 ps for $P=1.5$. For practical applications, the dashed line
represents $42 \Delta t$ during which the oscillatory segregation
stays larger than one tenth of its maximal value. This is a valuable
result, if one wishes to compare the segregation duration to a
characteristic time of some process likely to be enhanced by
species segregation.

%%%%%%%%%%%%%%%%%%%%%%%%%%%%%%%%%%%%%%%%%%%%%%%%%%%%%%%%%%
\section{Application and discussion}
\label{secdiscuss}
%%%%%%%%%%%%%%%%%%%%%%%%%%%%%%%%%%%%%%%%%%%%%%%%%%%%%%%%%%
The above results should now be applied to real binary mixtures to
assess the importance of the phenomenon.  Rather than selecting
specific mixtures, we will try to cover a wide range of molecule sizes
by taking typical values for the other mixture parameters.

We first combine equations (\ref{cadim}), (\ref{csmoothwall}) and
(\ref{coscwall}) to obtain the segregation ratio at the bubble
wall
\begin{equation}
  \label{concratio}
  \frac{\cdim(0,\taunl)}{\cdimz} = \exp(\beta I)
  \left(1 + \frac{\beta}{\pe^{1/2}}G(\taunl) \right).
\end{equation}
Having practical applications in view, we are interested in the
average and peak concentrations at the bubble wall, so that in what
follows, we will calculate the two quantities:
\begin{subeqnarray}
  \slabel{concsm}
  \segratsm &=& \exp(\beta I), \\
  \slabel{concosc}
  \segratosc &=& 
  \exp(\beta I) \frac{\beta}{\pe^{1/2}} \gmin,
\end{subeqnarray}
where $\gmin<0$ is calculated from equation (\ref{defgappmin}). The
two quantities $\segratsm$ and $\segratosc$ should be interpreted as
follows: \addedthree{the first} is the average concentration at the
bubble wall and the second is the maximum algebraic variation of the
concentration around \addedthree{the average}, over an oscillation period.

Before specifying the mixture, it is worth recalling that $\beta$
depends on two physical properties (see equation
(\ref{defbetam}))\addedthree{. On one hand} the relative densities of
species A and the host liquid, \addedthree{on the other hand} the
molar weight of species A. The latter may vary in a much larger range
than the former, so that in what follows, we will study the
predictions of the model for a mixture of water with a heavier species
A of apparent density $\rho_A = {M_A}/{\bar{V}_A} = 2000$\kgpermcube,
and molecular weights $M_A$ ranging from 100 to $10^7$\dalton{}
\addedthree{(the symbol Da stands for ``Dalton'' and 1\dalton{}
  corresponds to a molar weight of 1\gpermol)}.

The temperature of the mixture is set to $T=298$ K.  Consistently with
the dilute mixture hypothesis detailed in appendix \ref{annlinear},
the density $\rho$ of the mixture is approximated \addedthree{by the
  density of pure water}.  To calculate the oscillatory part under the
same conditions, the additional data of the diffusion coefficient is
needed. Since we consider a set of species which molecular weight
varies \addedthree{over} a very wide range, the influence of the
\addedthree{molecular} size on the diffusion coefficient $D$ should be
taken into account.  Following the Stokes-Einstein theory, the
diffusion coefficient can be expressed as
\begin{equation}
  \label{diffstokeseinstein}
  D = \frac{k_BT}{6\pi\mu R_A},
\end{equation}
where $k_B$ is the Boltzmann constant and $R_A$ the hydrodynamic
radius, estimated from the molecular weight and apparent density of
species A by
\begin{equation}
  \navogadro \frac{4}{3}\pi R_A^3 \rho_A = M_A,
\end{equation}
where $\navogadro$ is the Avogadro number. Under these conditions, the
parameter $\diml{\beta}_m$ defined by (\ref{defbetam}) ranges from
$-10^{-5}$ to $-1$\secpermcar, the hydrodynamic radius from 0.27 to
12.5\nm{}, and the diffusion coefficient from
$7.9\times10^{-10}$ to $1.7\times10^{-11}$\mcarpersec.

We consider the case of a 4\microns{} argon bubble excited at $f=$ 26.5
kHz. The corresponding P\'eclet number for the above conditions ranges
from 3\,360 to 15\,600, which justifies a posteriori the asymptotic
expansions in terms of $\pe^{-1/2}$, and the non-dimensional parameter
$\beta$ ranges from $-4.6\times10^{-6}$ to $-4.6\times10^{-1}$.

\begin{figure}
  \centerline{\includegraphics[width=0.9\linewidth]{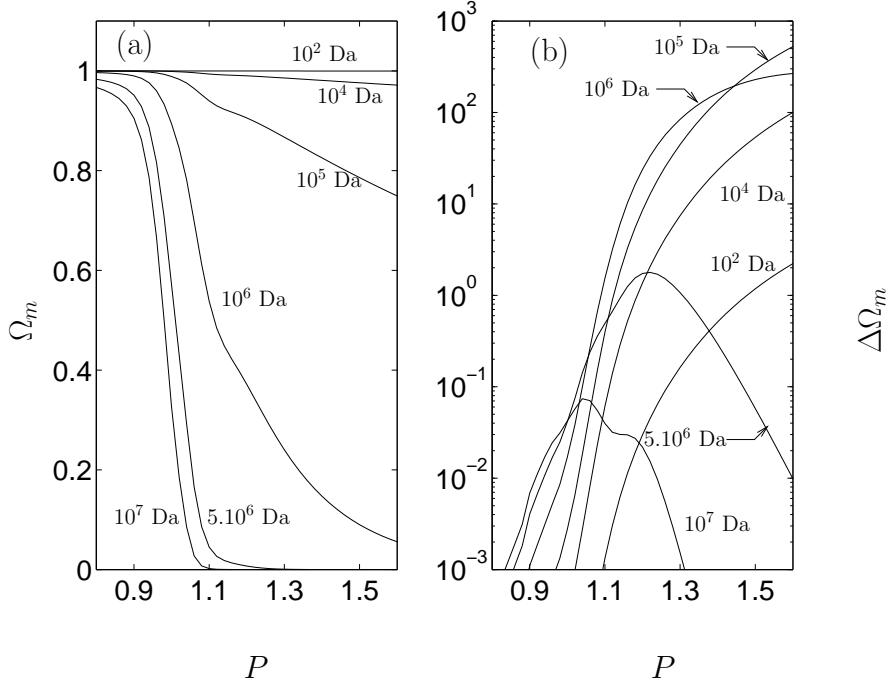}}
  \caption{Segregation ratio in a mixture of water with molecules of
    apparent density 2000\kgpermcube, of molecular weight $M_A$
    ranging from 100 to $10^7$\dalton{} around a 4\microns{} argon
    bubble driven at 26.5 kHz: (\textit{a}) smooth segregation ratio
    defined by equation (\ref{concsm}), (\textit{b}) oscillatory peak
    segregation ratio defined by equation (\ref{concosc}).}
  \label{figsolidparticles}
\end{figure}

The \addedthree{order} of magnitude of the average bubble wall concentration of
such molecules is \addedthree{shown} in figure
\ref{figsolidparticles}(\textit{a}): the $\segratsm$ curve for the
smallest molecules ($M_A=100$\dalton) remains indistinguishable from 1
even for high driving pressure so that the mixture is unsegregated on
average. As the weight of the molecules increases, their average
depletion at the bubble wall becomes increasingly high for a given
driving pressure. A nearly total depletion of the heaviest molecules
($M_A = 5\times 10^6$ and $10^7$\dalton) can even be observed for
driving pressures slightly above the Blake threshold.

The amplitude of the oscillatory concentration variation $\segratosc$
is \addedthree{shown} in figure
\ref{figsolidparticles}(\textit{b})\addedthree{, where} it can be seen
that the smallest molecule is already over-concentrated by \addedall{a
  factor of 2} at $P=1.5$. As $M_A$ increases, $\segratosc$ first
increases, and then decreases again for very large molecules.  This
illustrates the opposite effects of the two factors $\exp(\beta I)$
and $\beta\pe^{-1/2}\gmin$ in equation (\ref{concosc})\addedthree{.
  For} the smallest molecules, the increase of $|\gmin|$ with $P$
dominates over the decrease of $\exp(\beta I)$, so that $\segratosc$
globally increases with driving pressure, up to nearly 500 for
$M_A=100000$\dalton{} and $P=1.6$.  For larger molecules, the opposite
occurs, so that the peak value $\segratosc$ becomes increasingly
masked by the strong average depletion $\segratsm$ and
\addedall{hardly departs from 0} for $M_A=10^7$\dalton.

Thus, it is seen that both average depletion and peak periodic
over-concentration at the bubble wall compete, depending on the
driving level and the molecule sizes. This \addedthree{suggests} that
molecules or nano-particles that \addedthree{could} undergo some growth
or agglomeration process would be periodically concentrated against
the bubble wall as long as they are sufficiently small, but would be
held far from the bubble \addedthree{on average}, as they reach some critical size.
This may have some strong consequences on polymerization or
nano-particles agglomeration processes for example.

The present results may also help to understand the positive effect of
acoustic cavitation on \addedthree{crystal} nucleation from a solute
\cite[see for example][for potassium sulfate
crystallisation]{lyczkoespit2002}.  Homogeneous nucleation of crystals
in liquids is a first-order phase transition, which occurs as the
solute concentration exceeds the saturation concentration. There is a
fairly general agreement on the so-called classical nucleation theory
\cite[][]{Kashiev}, which states that in a metastable solution, the
nucleation process occurs through progressive accumulation of solute
molecules, forming multi-mers called ``clusters'', up to a critical
radius called nucleus, from which a solid crystal is then free to
grow. There is indeed experimental evidence of the existence of such
clusters and their stratification under gravity has been observed in a
sedimentation column \cite[][]{mullinleci69,larsongarside86}.
Although stated differently by these authors, the process invoked to
explain \addedthree{cluster} sedimentation obeys the pressure
diffusion equation considered in this paper. One may therefore
reasonably conjecture that the pressure diffusion effect around an
oscillating bubble would also tend to segregate these clusters to a
much larger extent than gravity, in view of the respective
accelerations involved. The present conclusions show that this is
indeed the case, and predict that if the nucleating species is heavier
than the liquid, its clusters would be periodically pushed against the
bubble wall. There, since the collision probability varies with the
\addedthree{square of the concentration}, they could undergo more
frequent attachment events and create bigger clusters. Above a
critical size these clusters would then be held far from the bubble in
the liquid, as suggested by figure
\ref{figsolidparticles}(\textit{a}).

\addedtwo{It should} be added that the conclusions on the smooth effect
should be \addedthree{tempered} by considerations on the bubble
stability\addedthree{. The} smooth effect needs a very large number of
acoustic periods to build up, so that its potential appearance is
conditioned by the bubble stability on such a large timescale. If this
stability is well established in single-bubble experiments, there is
no definitive conclusion in multi-bubble fields. This issue has been
discussed recently by \cite{louisnardgomez2003}\addedthree{. Even} in the most
optimistic case, inertial bubbles would rapidly increase their size by
rectified diffusion up to the fragmentation threshold, in a time too
small for the smooth effect to build up completely. Partial build-up
remains however possible, and may yield a noticeable smooth effect on
the largest molecules.

\addedtwo{Finally, despite the compressibility effects were
  arbitrarily neglected to reduce the mathematical complexity of the
  problem, it is nevertheless an important issue.  A spherical
  shock-wave can build up when the bubble rebounds after the collapse,
  and the corresponding steepening of the pressure profile may
  therefore enhance the oscillatory effect. Such spherical shocks are
  non-monotonic \cite[see for example figures 7 and 8 of][]{fujikawa}
  so that just after the collapse, the heaviest species would be very
  concentrated in a thin layer of fluid surrounding the shock, and
  would travel with the shock. An important consequence of this
  feature is that the excess concentration would not remain located
  near the bubble wall, but would be transported toward the bulk
  liquid. In summary, shock-waves would not only enhance the
  oscillatory effect, but they may also extend its influence to a
  larger spatial region.}

%%%%%%%%%%%%%%%%%%%%%%%%%%%%%%%%%%%%%%%%%%%%%%%%%%%%%%%%%%
\section{Conclusion}
\label{secconclusion}
%%%%%%%%%%%%%%%%%%%%%%%%%%%%%%%%%%%%%%%%%%%%%%%%%%%%%%%%%%
We have proposed an analytic method to solve the general problem of
pressure-gradient forced diffusion of two non-volatile species around
a bubble oscillating radially in the mixture. The method yields the
concentration field in the mixture around the bubble in two parts:
\addedthree{a smooth part}, building over a number of acoustic periods
of order of the P\'eclet number $\pe$ and asymptotically constant in
time, and an \addedthree{oscillatory part}. Both expressions are fully
analytic and can be easily calculated for a given bubble dynamics.

In the case of inertial cavitation, the oscillatory effect results in
a large excess concentration of the heaviest species at the bubble
wall at each bubble collapse. This excess is noticeable even for small
molecules, and relaxes with a characteristic time which is more than
one order of magnitude larger than the characteristic duration of the
bubble rebound.  Conversely, the smooth effect pushes the heaviest
species far from the bubble. It remains unimportant for small
molecules, even for strong driving pressures, but may \addedthree{almost deplete}
the bubble wall \addedthree{of} large molecules.  Both smooth and oscillatory
effects increase with driving pressure.  The smooth effect increases
with the frequency of the driving. The oscillatory effect increases
with frequency for small driving pressure but conversely, decreases
with frequency in the inertial regime.

For large molecules or nano-particles around an inertial bubble, the
two smooth and oscillatory effects compete: the oscillatory effect
dominates for the smallest molecules, while the smooth one is
prominent for the largest ones. This has strong implications for any
physico-chemical process involving molecules or particles undergoing a
growing or agglomeration process, and suggests that species smaller
than a given size would be periodically pushed and concentrated near
the bubble wall, while the largest ones are in average held far from
the bubble.  Polymerization, agglomeration or cluster formation in
crystal nucleation fall in this specific case and this behaviour may
be partly responsible for the reported enhancement of nucleation by
cavitation.

%%%%%%%%%%%%%%%%%%%%%%%%%%%%%%%%%%%%%%%%%%%%%%%%%%%%%%%%%%
\begin{acknowledgments}
  This work is supported by an ECOS-South collaboration program
  between France and Chile under grant number C03E05.
\end{acknowledgments}
%%%%%%%%%%%%%%%%%%%%%%%%%%%%%%%%%%%%%%%%%%%%%%%%%%%%%%%%%%

%%%%%%%%%%%%%%%%%%%%%%%%%%%%%%%%%%%%%%%%%%%%%%%%%%%%%%%%%%
\appendix
\section{Linearization of the convection--diffusion equation}
\label{annlinear}
%%%%%%%%%%%%%%%%%%%%%%%%%%%%%%%%%%%%%%%%%%%%%%%%%%%%%%%%%%
We assume an ideal mixture of two liquids, so that volume is
additive. Under these conditions, the mean density of the mixture is
\begin{equation}
  \label{defrhomoy}
  \rho = \frac{x_AM_A+x_BM_B}{x_A\bar{V}_A+x_B\bar{V}_B},
\end{equation}
where $x_i$, $M_i$ and $\bar{V}_i$ are the mole fraction, molecular
weight and partial molal volume of species i, respectively. Using
the relation $x_i = M\omega_i/M_i$ between mole and mass fraction, one
readily obtain
\begin{equation}
  \label{defrhomoybis}
  \rho = \frac{1}{\omega_A\bar{v}_A + \omega_B\bar{v}_B},
\end{equation}
where the notation $\bar{v}_i = \bar{V}_i/M_i$ has been used.
The mean molecular weight of the mixture is defined by
\begin{equation}
  \label{defMmoy}
  \frac{1}{M} = \frac{\oa}{M_A} + \frac{\omega_B}{M_B}.
\end{equation}
Replacing $\omega_B$ by $1-\oa$, \addedall{we can express the density of the
solution by}
\begin{equation*}
  \rho = \frac{1}{\vb} \left[1+\alpha_1\oa \right]^{-1},
\end{equation*}
and \addedall{the two contributions to diffusion} in equation (\ref{conservomgen})
become
\begin{subequations}
  \label{diffusionterms}
  \begin{equation}
    \slabel{fickterm}
    \rho\bnabla\oa = \frac{1}{\vb}
    \left[1+\alpha_1\oa \right]^{-1}\bnabla\oa,
  \end{equation}
  \begin{equation}
    \slabel{pressureterm}
    \frac{M_A M_B}{M\cgp T}
    \oa\left(\frac{\bar{V}_A}{M_A}-\frac{1}{\rho} \right)\bnabla p
    = \frac{M_A}{\cgp T}\alpha_1\oa
    \frac{(1-\oa)(1+\alpha_2\oa)}{1+\alpha_1\oa} \bnabla p,
  \end{equation}
\end{subequations}
where parameters $\alpha_1$ and $\alpha_2$ are defined by:
\begin{subequations}
  \label{defalphas}
  \begin{equation}
    \slabel{defalpha1}
    \alpha_1 = \frac{\bar{v}_A}{\bar{v}_B} - 1,
  \end{equation}
  \begin{equation}
    \slabel{defalpha2}
    \alpha_2 = \frac{M_B}{M_A} - 1.
  \end{equation}
\end{subequations}
Therefore it is seen that if we neglect terms of order $O(\oa^2)$,
$O(\alpha_1^2 \oa^2)$ and $O(\alpha_2^2 \oa^2)$, equation
(\ref{conservomgen}) becomes
\begin{equation}
  \label{rhsconservomgensimp}
  \frac{1}{\bar{v}_B}\left(\dsurd{\omega_A}{t} + \vv{v}\bcdot\bnabla\omega_A \right) = 
  D \frac{1}{\bar{v}_B}\bnabla\bcdot
    \left[\bnabla\oa + 
      \frac{M_A}{\cgp T}\oa\left(\bar{v}_A-\bar{v}_B \right) 
      \bnabla{p}\right]
\end{equation}
which is equation (\ref{conservomdilue}).

%%%%%%%%%%%%%%%%%%%%%%%%%%%%%%%%%%%%%%%%%%%%%%%%%%%%%%%%%%
\section{Solution of the oscillatory problem and splitting}
\label{annosc}
%%%%%%%%%%%%%%%%%%%%%%%%%%%%%%%%%%%%%%%%%%%%%%%%%%%%%%%%%%
Each member of the hierarchy of oscillatory problems may be expressed
as a non-homogeneous diffusion partial differential equation
\begin{equation}
  \label{eqnonhomog}
  \dsurd{\cosc_i}{\taunl}-\ddesurd{\cosc_i}{s} = 
  \gfi \left( R(\taunl),s,\cosc_0,\dots,\cosc_{i-1} \right),
\end{equation}
with a Neumann inhomogeneous boundary condition of the form:
\begin{equation}
  \label{clnonhomog}
  \dsurd{\cosc_i}{s}\left(s=0,\taunl \right) = 
  \gbi \left(R(\taunl),\cosc_{i-1} \right).
\end{equation}
We treat the problem in the manner of \cite{fyrillasszeri2}, with the
difference in that here we have a Neumann condition rather than a
Dirichlet one.

The oscillatory solution $\cosc_i$ should vanish for $s \rightarrow
+\infty$ since the outer solution of the boundary layer problem is
imposed to be identically 0. The asymptotic oscillatory solutions
$\coscas$ at any order have $T$ periodicity in the $\tau$ variable,
and therefore $\pernl = \taunl(T)$ periodicity in the $\taunl$
variable. Thus the functions $\gfi$ y $\gbi$ are also periodic in
$\taunl$ and we expand them in Fourier series, as well as $\coscas$.
Setting $\omega_m = 2m\pi/\pernl$, we get
\begin{eqnarray*}
  \coscas_i(s,\taunl) &=&
  \sum_{m=-\infty}^{m=+\infty} c_m(s)\exp(\icomp\omega_m\taunl), \\
  \gfi(s,\taunl) &=&
  \sum_{m=-\infty}^{m=+\infty} f_m(s) \exp(\icomp\omega_m\taunl), \\
  \gbi(\taunl) &=&
  \sum_{m=-\infty}^{m=+\infty} b_m \exp(\icomp\omega_m\taunl).
\end{eqnarray*}
Substituting these series in the problem
(\ref{eqnonhomog})-(\ref{clnonhomog}), we obtain a set of differential
equations relating the coefficients of these series. For any $m\ne 0$,
we obtain
\begin{gather}
  \ds\frac{d^2 c_m(s)}{ds^2} - \icomp\omega_m c_m(s) = - f_m(s)
  \label{eqcoefosc} \\
  \ds\frac{d c_m(s)}{ds}(s=0) = b_m
  \label{bdcoefosc} 
\end{gather}
The general solution of equation
(\ref{eqcoefosc}) vanishing for $s \rightarrow\infty$ is
\begin{equation}
  \label{solosc}
  c_m(s) = A_m\exp\left(-k_m s \right) 
  - \frac{1}{k_m}
  \int_s^{\infty}
  f_m(s') \sinh \left[ k_m (s-s') \right] \:\ddint s',
\end{equation}
with
\[
k_m = (1+\addedall{\sgnm}\icomp)\left(\frac{\addedall{|}\omega_m\addedall{|}}{2} \right)^{\shalf} 
= \left(\frac{2\addedall{|}m\addedall{|}\pi}{\pernl} \right)^{\shalf}
e^{\icomp\addedall{\sgnm}\pi/4},
\]
\addedall{where $\sgnm=\sgn{(m)}$.}
The boundary condition at $s=0$ (\ref{bdcoefosc}) yields the following
expression for $A_m$:
\begin{equation}
  \label{defAm}
  A_m = -\frac{b_m}{k_m} - \frac{1}{k_m}
  \int_0^\infty f_m(s') \cosh (k_m s') \:\ddint s'.
\end{equation}
The zeroth-order harmonics differential equation ($m=0$) takes a
different form:
\begin{equation*}
  \frac{d^2 c_0(s)}{ds^2}  = - f_0(s),
\end{equation*}
with the associated boundary condition
\begin{equation*}
  \label{coefconst}
  \frac{d c_0(s)}{ds}(s=0) = b_0.
\end{equation*}
The solution vanishing for $s=\infty$ is 
\[
c_0(s) = \int_\infty^s \int_{s'}^\infty f_0(s'')
\:\ddint s'' \:\:\ddint s'.
\]
Applying the Neumann boundary condition at $s=0$  yields:
\[
b_0 = \int_0^\infty f_0(s') \:\ddint s',
\]
and recognizing that $b_0 = \left< \gbi\right>_{\taunl}$ and
$f_0(s) = \left< \gfi(s)\right>_{\taunl}$, the separation condition
finally reads 
\begin{equation}
\label{splitcond}
  \left< \gbi\right>_{\taunl} = \int_0^{+\infty} \left<
    \gfi(s)\right>_{\taunl} \:\ddint s.
\end{equation}
Finally, in the special case where $F^i$ is identically zero, which is
the case in the present paper for $i=1$, the separation condition just
implies that $\left< \gbi\right>_{\taunl}$ should be 0, and the
oscillatory field reads in this case:
\begin{equation}
  \label{coscnofi}
  \coscas_i(0,\taunl)  = - \left(\frac{\pernl}{2\pi} \right)^{\shalf}
  \sum_{%
    \substack{
      m = -\infty \\
      m \neq 0}}^{m=+\infty} \frac{b_m}{\addedall{|}m\addedall{|}^{1/2}}
  \exp\left[\icomp\left( 2\pi m \frac{\taunl}{\pernl} -\addedall{\sgnm}\frac{\pi}{4}\right)
    -(\addedall{\sgnm}\icomp+1)
    \left(\frac{\addedall{|}m\addedall{|}\pi}{\pernl}\right)^\shalf s \right].
\end{equation}
It can be noted that the presence of both $m^{1/2}$ in the
denominator and the $\pi/4$ phase lag recalls the fact that
$\coscas_i(0,\taunl)$ is the half-order integral of $\gbi(\taunl)$ as
could be proved directly by solving problem (\ref{eqnonhomog}),
(\ref{clnonhomog}) by Laplace transforms.   

% %%%%%%%%%%%%%%%%%%%%%%%%%%%%%%%%%%%%%%%%%%%%%%%%%%%%%%%%%%
% \section{Numerical estimation of the smooth asymptotic concentration field}
% \label{anncalcsm0}
% %%%%%%%%%%%%%%%%%%%%%%%%%%%%%%%%%%%%%%%%%%%%%%%%%%%%%%%%%%
%%%%%%%%%%%%%%%%%%%%%%%%%%%%%%%%%%%%%%%%%%%%%%%%%%%%%%%%%%
\section{Solution of the second-order oscillatory problem}
\label{annoscordre2}
%%%%%%%%%%%%%%%%%%%%%%%%%%%%%%%%%%%%%%%%%%%%%%%%%%%%%%%%%%
The second-order oscillatory problem reads
\begin{subequations}
  \label{pbmosc2}
  \begin{equation}
    \label{eqosc2}
      \ds  \dsurd{\cosc_2}{\taunl} = \ddesurd{\cosc_2}{s} + 
      \dsurd{}{s}\left[\abis_1 s\dsurd{\cosc_1}{s}
        + \beta\bbis_0\cosc_1\right],
  \end{equation}
  \begin{equation}
    \label{clbulosc2}
      \ds \dsurd{\cosc_2}{s}(0,\taunl) + \beta H(\taunl) \cosc_1(0,\taunl)= 
      -\gsep_1 -  \beta H(\taunl)\csm_1(0,\tau),
  \end{equation}
  \begin{equation}
    \label{clinfosc2}
    \cosc_2(s\rightarrow\infty) = 0.
  \end{equation}
\end{subequations}
The splitting condition (\ref{splitcond}) reads therefore:
\begin{multline}
  \label{ccu2}
  \left< - \beta H(\taunl) \cosc_1(0,\taunl) -
    \gsep_1 - \beta H(\taunl) \csm_1(0,\tau) \right>_{\taunl}\\
=  \left<\int_0^{\infty}
      \dsurd{}{s}\left(\abis_1 s \dsurd{\cosc_1}{s}+\beta \bbis_0\cosc_1 \right)
      \:\ddint s
  \right>_{\taunl}.
\end{multline}
Using equations (\ref{relABH}) and (\ref{clinfosc1}), the integral in
the right-hand side of equation (\ref{ccu2}) can also be written
\begin{equation}
  \label{intrhs}
  \lim_{s\rightarrow\infty}  \left(
    \abis_1 s \dsurd{\cosc_1}{s}(s,\taunl)
  \right)
  -\beta H(\taunl) \cosc_1(0,\taunl).
  % + \int_0^{\infty}\beta \fsep_0\bbis_1 \:\ddint s
\end{equation}
It can be seen from equation (\ref{coscresult}) that the first term of
expression (\ref{intrhs}) is zero, so that (\ref{ccu2}) becomes
finally
\begin{equation}
  \label{defgsep1ann}
  \gsep_1 = - \beta
    \left< 
      H(\taunl)\left[
        \csm_1(0,\tau)
      \right] 
    \right>_{\taunl}.
\end{equation}

%%%%%%%%%%%%%%%%%%%%%%%%%%%%%%%%%%%%%%%%%%%%%%%%%%%%%%%%%%
\section{Solution of the smooth problem}
\label{annsmooth}
%%%%%%%%%%%%%%%%%%%%%%%%%%%%%%%%%%%%%%%%%%%%%%%%%%%%%%%%%%
The second and third order smooth equations are
\begin{subeqnarray}
  \label{eqsmooth23}
  \slabel{eqsmooth2}
  \dsurd{\csm_2}{\tau} &=& - \dsurd{\csm_0}{\lambda} +
  \dsurd{}{\sigma}
  \left[
    A(\sigma,\tau) \dsurd{\csm_0}{\sigma} + 
    \beta B(\sigma,\tau) (\csm_0+1)
  \right],
  \\
  \slabel{eqsmooth3}
  \dsurd{\csm_3}{\tau} &=& - \dsurd{\csm_1}{\lambda} +
  \dsurd{}{\sigma}
  \left[
    A(\sigma,\tau) \dsurd{\csm_1}{\sigma} + 
    \beta B(\sigma,\tau) \csm_1
  \right].
\end{subeqnarray}
The expansion (\ref{devcsm}) must be uniformly valid and therefore
should not contain secular terms increasing unbounded when
$\tau\rightarrow\infty$. This non-secular behaviour will be satisfied
by $\csm_2$ and $\csm_3$ only if the right-hand sides of equations
(\ref{eqsmooth23}\,\textit{a,b}) have zero $\tau$-averages. Therefore,
$\csm_0$ and $\csm_1$ should fulfill the respective non-secularity
conditions
\begin{subequations}
  \label{eqsm01}
  \begin{equation}
    \slabel{eqsm0}
    \dsurd{\csm_0}{\lambda} = 
    \dsurd{}{\sigma}
    \left[
      \left<A(\sigma,\tau) \right>_{\tau} \dsurd{\csm_0}{\sigma} + 
      \beta \left<B(\sigma,\tau) \right>_{\tau} (\csm_0 + 1) \right],
  \end{equation}
  \begin{equation}
    \slabel{eqsm1}
    \dsurd{\csm_1}{\lambda} = 
    \dsurd{}{\sigma}
    \left[
      \left<A(\sigma,\tau) \right>_{\tau} \dsurd{\csm_1}{\sigma} + 
      \beta \left<B(\sigma,\tau) \right>_{\tau} \csm_1 \right],  
  \end{equation}
\end{subequations}
where the independence of $\csm_0$ and $\csm_1$ on $\tau$ has been
used. 

The associated boundary conditions at the bubble wall, equation
(\ref{clsmooth}), can be obtained from expressions (\ref{defgsep0})
and (\ref{defgsep1}) of the separation constants $\gsep_0$ and
$\gsep_1$.  Further using the independence of $\csm_0$ and $\csm_1$ on
the fast variable $\tau$, these boundary conditions read
\begin{subequations}
  \label{clsmprov01}
  \begin{equation}
    \label{clsm0prov}
    \dsurd{\csm_0}{\sigma}(0,\lambda) 
    + \beta \left< H(\taunl) \right>_{\taunl}
    \left[ \csm_0(0,\lambda) +1 \right] = 0,    
  \end{equation}
  \begin{equation}
    \label{clsm1prov}
    \dsurd{\csm_1}{\sigma}(0,\lambda) 
    + \beta \left< H(\taunl) \right>_{\taunl}
    \csm_1(0,\lambda) = 0.    
  \end{equation}
\end{subequations}
Moreover, it can be noticed that, from the definition (\ref{defH}) of
$H$, the nonlinear average $\left< H(\taunl) \right>_{\taunl}$ also reads
\[
  \label{valhmoy}
  \left<H \right>_{\taunl} 
  = \left<\frac{B(0,\tau)}{A(0,\tau)} \right>_{\taunl}
  = \frac{
    \left<\vqt B(0,\tau)/A(0,\tau) \right>_\tau
    }{\left<\vqt\right>_\tau}
  =\frac{\left<B(0,\tau)
    \right>_{\tau}}{\left<A(0,\tau) \right>_\tau},
\]
since $A(0,\tau)=\vqt$, so that the zeroth- and first-order boundary
conditions (\ref{clsmprov01}\,\textit{a,b})  at the bubble wall may also be written
\begin{subequations}
  \label{clsm01}
  \begin{equation}
    \slabel{clsm0}
    \left<A(0,\tau) \right>_{\tau} 
    \dsurd{\csm_0}{\sigma}(0,\lambda) 
    + \beta \left<B(0,\tau) \right>_{\tau}
    \left[ \csm_0(0,\lambda) +1 \right] = 0,
  \end{equation}
  \begin{equation}
    \slabel{clsm1}
    \left<A(0,\tau) \right>_{\tau} 
    \dsurd{\csm_1}{\sigma}(0,\lambda) 
    + \beta \left<B(0,\tau) \right>_{\tau}
    \csm_1(0,\lambda) = 0.
  \end{equation}
\end{subequations}
We now seek the asymptotic solutions $\cii$ for $i=0,1$ of equations
(\ref{eqsm01}\,\textit{a,b}), by setting $\partial\cii/\partial\lambda = 0$ for
$i=0,1$ in these equations and integrating once with respect to
$\sigma$.  Making use of boundary conditions (\ref{clsm01}\,\textit{a,b}), this
integration yields
\begin{subequations}
  \label{eqsminteg01}
  \begin{equation}
    \slabel{eqsminteg0}
    \left<A(\sigma,\tau) \right>_{\tau} 
    \dsurd{\czi}{\sigma}
    + \beta \left<B(\sigma,\tau) \right>_{\tau}
    \left( \czi +1 \right) = 0,
  \end{equation}
  \begin{equation}
    \slabel{eqsminteg1}
    \left<A(\sigma,\tau) \right>_{\tau} 
    \dsurd{\cui}{\sigma}
    + \beta \left<B(\sigma,\tau) \right>_{\tau}
    \cui = 0.
  \end{equation}
\end{subequations}
Now using the condition at infinity (\ref{clinfsm}), the zeroth-order
equation (\ref{eqsminteg0}) can be integrated as
\begin{equation}
  \label{solsmoothfinalann}
  \czi(\sigma) = \exp\left[\beta\int_\sigma^\infty
%    \left<\ds\frac{4}{9}\ds\frac{\vp^2}{(3\sigma'+V)} \right>_\tau
%    \frac{\ddint\sigma'}{\left<(3\sigma'+V)^{4/3}\right>_\tau}
    \frac{\left<B(\sigma',\tau) \right>_{\tau}}
    {\left<A(\sigma',\tau) \right>_{\tau}}\:\ddint\sigma'
    \right] - 1.
\end{equation}
Besides, integration of the first-order equation (\ref{eqsminteg1})
can only yield the null solution $\cui=0$ in order to fulfill the
condition at infinity, equation (\ref{clinfsm}). It does not imply
however that $\csm_1(\sigma,\lambda)$ is zero for finite $\lambda$,
but just states that its asymptotic limit for
$\lambda\rightarrow\infty$ is zero.

The physical meaning of the asymptotic smooth solution may be
understood by at equations (\ref{eqsminteg01}\,\textit{a,b}).
\addedthree{The} average pressure diffusion flux (the $B$ term) is
exactly balanced by the average Fick diffusion flux (the $A$ term),
and therefore the smooth concentration field stays constant. The
unsteady term in the smooth equations (\ref{eqsm01}\,\textit{a,b})
represents the transitory non-equilibrium between the two average
diffusion processes.

%%%%%%%%%%%%%%%%%%%%%%%%%%%%%%%%%%%%%%%%%%%%%%%%%%%%%%%%%%
\section{Numerical estimation of the oscillatory asymptotic
  concentration field}
\label{annpeak}
\begin{figure}
  \centerline{\includegraphics[width=0.9\linewidth]{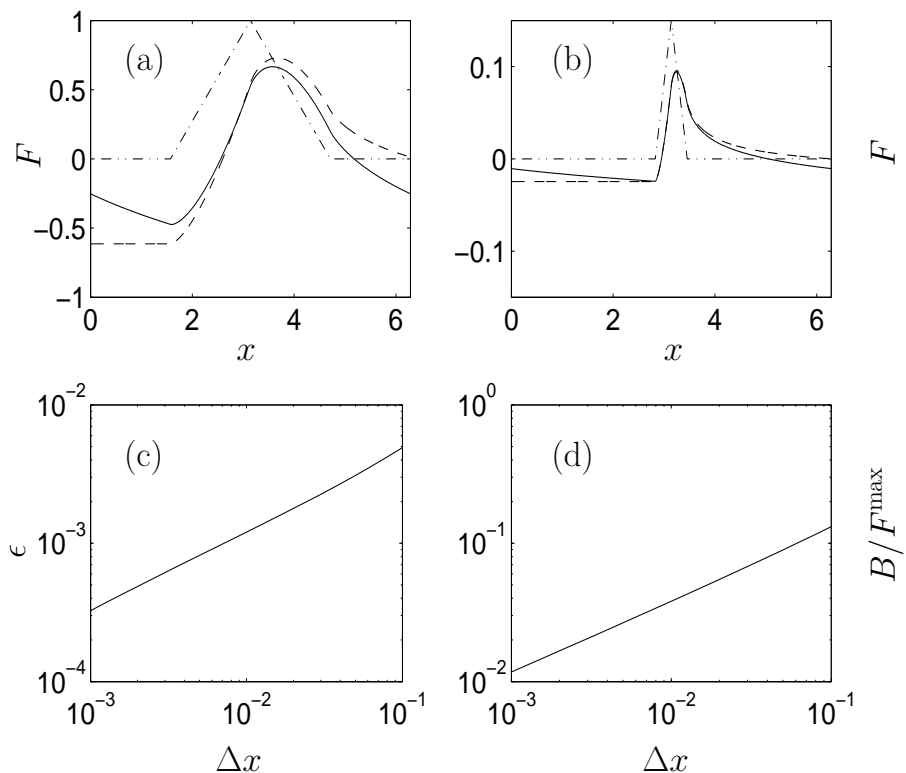}}
  \caption{(\textit{a}) and (\textit{b}): comparison of function $F$ calculated numerically from
    equation (\ref{trueF}) (solid line) with function $F$ calculated by
    approximation (\ref{fapprox}) (dashed line). The dash-dotted line
    recalls the shape of the tooth approximation (\ref{htooth}) of function $H$.
    Figure (\textit{a}) is obtained with $\dx=\pi/2$ and figure
    (\textit{b}) with $\dx=\pi/10$. (\textit{c}) Relative error on the maximum
    value of $F$ calculated from (\ref{fapprox}), as $\dx$ is varied.
    (\textit{d}) Ratio $B/\fmax$ as $\dx$ is varied.  }
  \label{figFapprox}
\end{figure}

%%%%%%%%%%%%%%%%%%%%%%%%%%%%%%%%%%%%%%%%%%%%%%%%%%%%%%%%%%
We first set $x=2\pi\taunl/\pernl$, $\xmin = 2\pi\taunlmin/\pernl$ and
$\dx = 2\pi\Delta\taunl/\pernl$ . In the new variable $x$, the
tooth-approximation (\ref{hfit}) can be written
\begin{equation}
  \label{htooth}
  H(x) \simeq \left\{
    \begin{array}{ll}
      \hmin
      \left(1-\left|\displaystyle\frac{x-\xmin}{\Delta x} \right|
      \right),
      &  x\in [\xmin- \Delta x, \xmin+ \Delta x ],
      \\
      0, & \text{elsewhere},
    \end{array}
    \right.
\end{equation}
with $\hmin<0$. This function can be Fourier-expanded as
\begin{equation}
  \label{hfourier}
  H(x) = \frac{H_m \Delta x}{2\pi} + \frac{2}{\pi\Delta x}\hmin
  \sum_{%
    \substack{
      m = -\infty \\
      m \neq 0}}^{m=+\infty}\frac{1}{m^2}
\sin^2\left(\frac{m\Delta x}{2} \right) \exp\left[\icomp m(x- \xmin) \right],
\end{equation}
and function $\gser$ defined by (\ref{defgser}) can therefore be
approximated as
\begin{equation}
  \label{gappann}
  \gapp(\taunl) = \left(\frac{\pernl}{2\pi}
  \right)^{1/2}\hmin\frac{4}{\pi\Delta x} F(x),
\end{equation}
with
\begin{equation}
  \label{trueF}
  F(x) = \sum_{m= 1}^{m=+\infty}\frac{1}{m^{5/2}} 
  \sin^2 \left(\frac{m\Delta x}{2} \right) 
  \cos\left[m(x-\xmin) -\frac{\pi}{4}\right],
\end{equation}
which is equation (\ref{defgappser}). 
In order to get an estimate of the maximum of $F(x)$, we first
reformulate it as:
\begin{equation}
  \label{devs}
  F(x) = C(x) + S(x),
\end{equation}
with
\begin{subequations}
\label{defserieF}
\begin{multline}
  \label{defseriecos}
  C(x) = \frac{1}{4\sqrt{2}}\sum_{m=1}^{+\infty} \frac{1}{m^{5/2}} 
  [2\cos m(x-\xmin)\\
  -\cos m (x-\xmin+\dx)-\cos m (x-\xmin-\dx)]
\end{multline}
\begin{multline}
  \label{defseriesin}
  S(x) = \frac{1}{4\sqrt{2}}\sum_{m=1}^{+\infty} \frac{1}{m^{5/2}} 
  [2\sin m(x-\xmin)\\
  -\sin m (x-\xmin+\dx)-\sin m (x-\xmin-\dx)]
\end{multline}
\end{subequations}
%
%Each member is a serie of the type $\sum_{m=1}^{+\infty}m^{-5/2} \cos
%m X$ or $\sum_{m=1}^{+\infty}m^{-5/2} \sin m X$. 
Let's set:
\begin{equation}
  \label{cosplussin}
  Z(X) =  \sum_{m=1}^{+\infty}m^{-1/2} (\cos mX+\sin mX)
\end{equation}
It can be seen that differentiating (\ref{devs}), (\ref{defserieF}) twice,
$F''(x)$ is the sum of three series of the form (\ref{cosplussin}):
\begin{equation}
\label{fsecfoncdeg}
F''(x) =   \frac{1}{4\sqrt 2} \left[
  Z(x-\xmin-\dx) + Z(x-\xmin+\dx) - 2 Z(x-\xmin)
\right]
\end{equation}
A theorem by \cite{zygmund} states that
\begin{subequations}
  \label{thzygmund}
  \begin{equation}
    \label{thzygcos}
    \sum_{m=1}^{+\infty}m^{-\beta} \cos mX \underset{X\rightarrow 0}{\simeq}
    \left|X\right|^{\beta-1}\Gamma(1-\beta) \sin \pi\frac{\beta}{2}    
  \end{equation}
  \begin{equation}
    \label{thzygsin}
    \sum_{m=1}^{+\infty}m^{-\beta} \sin mX \underset{X\rightarrow 0}{\simeq}
    \sgn(X) \left|X\right|^{\beta-1}\Gamma(1-\beta) \cos \pi\frac{\beta}{2}
  \end{equation}
\end{subequations}
for any $\beta\in[0,1[$. Therefore, taking $\beta=1/2$, we get
\begin{equation}
  \label{thzygcossin}
  Z(X) \underset{X\rightarrow 0}{\simeq}
    \heav(X) \left|X\right|^{-1/2} \sqrt{2}\gamundemi
%    +\left|X\right|^{-1/2}\epsilon(X)
\end{equation}
where $\heav$ is the Heaviside function.  For small enough $\dx$, any
$x$ in the neighbourhood of $\xmin$ is also in the neighbourhood of
$\xmin-\dx$ and $\xmin+\dx$, so that, from (\ref{fsecfoncdeg}), we can
approximate $F''(x)$ as
\begin{multline}
  \label{fsecapprox}
  F''(x)\simeq \fapp''(x) \\= \frac{\gamundemi}{4}\left[
    \frac{\heav(x-\xmin-\dx)}{\left|x-\xmin-\dx \right|^{1/2}}
    + \frac{\heav(x-\xmin+\dx)}{\left|x-\xmin+\dx \right|^{1/2}}
    -2\frac{\heav(x-\xmin)}{\left|x-\xmin \right|^{1/2}}
  \right].
\end{multline}
Integrating twice yields
\begin{multline}
  \label{fapprox}
  F(x)\simeq \fapp(x) = \\
  \frac{\gamundemi}{3}\left[
    f(x-\xmin-\dx) + f(x-\xmin+\dx) -2f(x-\xmin)
  \right]
  +Ax+B, 
\end{multline}
where $f$ is defined by
\begin{equation*}
  f(X) = \heav(X)\left|X\right|^{3/2},
\end{equation*}
and $A$, $B$ are two integration constants. Clearly $A$ should be 0 to
avoid a spurious discontinuity of $F$ at $x=2n\pi$ and $B$ must be calculated
so that the approximation of $F$ has a zero average on $[0,2\pi]$, as
does the original function (\ref{trueF}). This condition yields 
\begin{equation*}
  B = \frac{\gamundemi}{15\pi}\left[2(2\pi-\xmin)^{5/2}
    - (2\pi-\xmin-\dx)^{5/2} - (2\pi-\xmin+\dx)^{5/2}
  \right].
\end{equation*}
It can be easily checked that $\fapp(x)$ has a maximum at $x=\xmin+\dx/3$
whose value is
\begin{equation}
  \label{Fmax}
  \fappmax(x) = \frac{2\gamundemi}{3\sqrt{3}} \dx^{3/2} + B.
\end{equation}
Owing to the approximation used to obtain (\ref{fsecapprox}), it is
clear that approximation (\ref{fapprox}) becomes better for smaller
$\dx$. Figure \ref{figFapprox}(\textit{a,b}) shows a comparison of the
calculated series (\ref{trueF}) (solid line) and its approximation by
(\ref{fapprox}) (dashed line) for $\xmin=\pi$ and $\dx=\pi/2$ (figure
\ref{figFapprox}\textit{a}) or $\dx=\pi/10$ (figure
\ref{figFapprox}\textit{b}). It is seen that for $\dx$ as large as
$\pi/2$ (\addedthree{in this case the peak spans}  over half of the interval), the
maximum of $F$ is still predicted with a relative error as low as 8
\%. For $\dx=\pi/10$ is reduced to 1.25 \%. Moreover, it can be noted
that the approximation of $F$ is not only good near $\xmin$, where it
should \addedthree{be}, but also over the whole interval $[0,2\pi]$.

The quality of the approximation of $\max_x{F}$ can be seen in figure
\ref{figFapprox}(\textit{c}), in which the relative error $\epsilon =
\left|\fmax-\fappmax\right|/\fmax$ is displayed as a function of
$\dx$: since for a typical inertial bubble, $\dx$ amounts to
$10^{-9}$, it is clear that the approximation given by equation
(\ref{Fmax}) is excellent.  We also draw the value of the constant $B$
relative to $\fappmax$ on figure \ref{figFapprox}(\textit{d}), which
shows clearly that $B$ can be easily neglected for $\dx$ smaller than
$10^{-2}$.

% \begin{figure}
%   \centerline{\includegraphics[width=0.9\linewidth]{figqualapprox}}
%   \caption{a: Relative error on the maximum value of $F$ calculated
%     from (\ref{fapprox}), as $\dx$ is varied. b: Ratio $B/\fmax$.  }
%   \label{figqualapprox}
% \end{figure}

Finally, it is of interest to know the characteristic relaxation time
of function $F$ after it has reached its maximum.  It can be shown
after some algebra that $F$ reaches a fraction of its maximum value
$\alpha \fmax$ after a time approximately equal to $27\Delta x/64
\alpha^2$. Applying this formula shows that $F$ is still equal to one
fifth of its maximum value after $10.5 \Delta x$, and to one tenth after
$42 \Delta x$. 

\bibliographystyle{jfm}
% \bibliography{abbrev,amortissement,bibliomecadef,bjerknes,champbulles,compressible,cristallisation,deuxbulles,dispersi,propagation,rectifie,surfacewaves,thermique,touseffets,twophase,unebulle,sonolum,hydrobulle,divers,mathappl,mespublis,numerical,sonochem,surfactant}
% %%%%%%%%%%%%%%%%%%%%%%%%%%%%%%%%%%%%%%%%%%%%%%%%%%%%%%%%%%%%%%%%%%%%%%

\end{document}